\documentclass[useAMS,usenatbib,usegraphicx]{mn2e}

\usepackage{deluxetable,amssymb}

\newcommand{\amm}{NH$_3$}
\newcommand{\dia}{N$_2$H$^+$}

\newcommand{\hcn}{HC$_7$N}

\newcommand{\kms}{km\,s$^{-1}$}
\newcommand{\cc}{cm$^{-3}$}
\newcommand{\vlsr}{$v_{\mbox{\tiny LSR}}$}
\newcommand{\mh}{$m_{\mbox{\tiny H}}$}

\date{Accepted 2013 September 3.  Received 2013 August 30; in original form 2013 June 27}

\title[Abundant cyanopolyynes as a probe of infall in Serpens South]{Abundant cyanopolyynes as a probe of infall in the Serpens South cluster-forming region}
\author[R. K. Friesen et al.]{R. K. Friesen,$^1$\thanks{E-mail:friesen@di.utoronto.ca} L. Medeiros,$^{2,3}$ S. Schnee,$^{3}$ T. L. Bourke,$^{4}$ J. Di Francesco,$^{5,6}$ 
\newauthor
R. Gutermuth,$^{7}$ P. C. Myers$^{4}$\\
$^{1}$Dunlap Institute for Astronomy and Astrophysics, University of Toronto, 50 St. George St., Toronto ON CANADA M5S 3H4 \\
$^{2}$University of California at Berkeley, Berkeley, CA 94720-3411 \\
$^{3}$National Radio Astronomy Observatory, 520 Edgemont Rd, Charlottesville VA 22903 \\
$^{4}$Radio and Geoastronomy Division, Harvard Smithsonian Center for Astrophysics, MS-42, Cambridge, MA, 02138, USA \\
$^{5}$National Research Council Canada, Radio Astronomy Program, 5071 West Saanich Rd, Victoria, BC, V9E 2E7, Canada \\
$^{6}$Department of Physics and Astronomy, University of Victoria, Victoria, BC, V8P 1A1, Canada \\
$^{7}$Department of Astronomy, University of Massachusetts, Amherst
} 
\begin{document}

\maketitle

\label{firstpage}

\begin{abstract}

We have detected bright \hcn\, $J = 21-20$ emission toward multiple locations in the Serpens South cluster-forming region using the K-Band Focal Plane Array at the Robert C. Byrd Green Bank Telescope. \hcn\, is seen primarily toward cold filamentary structures that have yet to form stars, largely avoiding the dense gas associated with small protostellar groups and the main central cluster of Serpens South. Where detected, the \hcn\, abundances are similar to those found in other nearby star-forming regions. Toward some \hcn\, `clumps', we find consistent variations in the line centroids relative to \amm\, (1,1) emission, as well as systematic increases in the \hcn\, non-thermal line widths, which we argue reveal infall motions onto dense filaments within Serpens South with minimum mass accretion rates of $M \sim 2-5$~M$_\odot$~Myr$^{-1}$. The relative abundance of \amm\, to \hcn\, suggests that the \hcn\, is tracing gas that has been at densities $n \sim 10^4$~\cc\, for timescales $t \lesssim 1-2 \times 10^5$~yr. Since \hcn\, emission peaks are rarely co-located with those of either \amm\, or continuum, it is likely that Serpens South is not particularly remarkable in its abundance of \hcn, but instead the serendipitous mapping of \hcn\, simultaneously with \amm\, has allowed us to detect \hcn\, at low abundances in regions where it otherwise may not have been looked for. This result extends the known star-forming regions containing significant \hcn\, emission from typically quiescent regions, like the Taurus molecular cloud, to more complex, active environments. 

\end{abstract}


\section{Introduction}
\label{sec:intro}

Cyanopolyynes are carbon-chain molecules of the form HC$_{2n+1}$N. At higher $n$, these molecules are among the longest and heaviest molecules found in the interstellar medium, and to date have been primarily seen toward several nearby, low-mass star-forming regions, and in the atmospheres of AGB stars \citep{winnewisser78}. For example, cyanohexatriyne (\hcn) has been detected toward several starless and protostellar cores within the Taurus molecular cloud \citep{kroto78, snell81, cernicharo86, olano88, sakai08} and more recently in Lupus \citep{sakai10}, Cepheus \citep{cordiner11} and Chamaeleon \citep{cordiner12}. The long cyanopolyynes are interesting for several reasons. They are organic species, and their abundance in star-forming regions raises questions about whether they are able to persist in the gas phase or on dust grains as star formation progresses, for example, in the disk around a protostar. While excited at temperatures and densities typical of star-forming regions, to form in high abundances cyanopolyynes require the presence of reactive carbon in the gas phase. In addition, the high abundances of long ($n \geq 3$) cyanopolyynes in regions like Taurus can be best explained by chemical reaction networks containing molecular anions like C$_6$H$^-$ \citep{cwalsh09}, which have only recently been discovered in the interstellar medium \citep{mccarthy06,gupta07,cordiner13}. Species like \hcn\, are thus expected to be abundant in star-forming regions where atomic carbon is not yet entirely locked up in CO, and to then deplete rapidly from the gas phase as the rates of destructive ion-molecule reactions dominate over formation reactions due to the depletion of atomic carbon. Consequently, measurements of the abundances of cyanopolyynes of several lengths, e.g., HC$_3$N, HC$_5$N, and \hcn, could be used as a probe of age in molecular cloud cores \citep{stahler84}. Here, we present the serendipitous discovery of multiple \hcn\, `clumps' within the young, cluster-forming Serpens South region within the Aquila rift. This is the first detection of \hcn\, in the Aquila rift, and also the first detection of \hcn\, in an active, cluster-forming environment. 

Serpens South is a star-forming region, associated in projection with the Aquila Rift, that contains a bright, young protostellar cluster (the Serpens South Cluster, or SSC) embedded within a hub-filament type formation of dense gas seen in absorption against the infrared background \citep{gutermuth08}. The cluster's large stellar density ($> 430$ stars\,pc$^{-2}$), relatively high cluster membership (48 young stellar objects, or YSOs, in the cluster, and 43 more along the filaments) and unusually high fraction (77\%) of Class I protostars among its YSOs are all evidence that star formation in the SSC has likely only very recently begun (e.g., within $(2 \pm 1) \times 10^5$\,yr), and that the star formation rate is high in the cluster core \citep[$\sim 90$\,M$_\odot$\,Myr$^{-1}$,][]{gutermuth08}. The dense gas filaments to the north and south have lower star formation efficiencies ($\sim 30$\,\% vs. $\sim 5$\,\%), with an average $SFR \sim 23$~M$_\odot$~Myr$^{-1}$~pc$^{-2}$ over the entire complex \citep{maury11}. Near-infrared polarimetry suggests the magnetic field in the region is coherent over the entire structure of dense gas, with field lines largely perpendicular to the main gas filament \citep{sugitani11}. Recently, \citet{kirk13} found evidence for accretion flows both onto and along one of the filaments at rates that are comparable to the current star formation rate in the central cluster. 

The distance to the SSC is currently debated. Serpens South lies within the highest extinction region along the Aquila Rift \citep[mid-cloud distance $d \sim 270$\,pc;][]{straizys03}, and is close in projection to the W40 OB association. The Serpens Main star-forming region lies 3\degr\, to the north. Local standard of rest (LSR) velocities are similar within all three objects. Recent VLBA parallax measurements, however, have shown that a YSO associated with Serpens Main is more distant than the Rift \citep[$d = 415 \pm 5$\,pc;][]{dzib10}, while distance estimates to W40 range from $d \sim 300$~pc to $d \sim 900$~pc [cite]. Through a comparison of the x-ray luminosity function of the young cluster powering the W40 HII region with the Orion cluster, \citet{kuhn10} put W40 at a best-fit distance of $\sim 600$\,pc, but do not rule out a smaller value. In contrast, the Herschel Gould Belt Survey assumed both the SSC and W40 are 260~pc distant \citep{andre10,bontemps10}. \citet{maury11} also argue that the SSC and W40 are at the same distance, citing their location within the same visual extinction feature presented by \citet{bontemps10}, and their similar velocities that span a continuum in \vlsr\, between 4\,\kms\, and 10\,\kms. The authors further suggest that the location of Serpens Main within a separate extinction feature, and its larger derived distance, may indicate that Serpens Main is behind the Aquila Rift (and hence the SSC and W40). In this work, we assume a distance of 260~pc to Serpens South to compare more readily with published results, but also discuss the implications of a larger distance on our analysis. 

We have performed observations of the \amm\, (1,1) and (2,2) inversion transitions, and simultaneously the \hcn\, $J=21-20$ rotational transition, toward the SSC and associated molecular gas using the 7-element K-band Focal Plane Array (KFPA) at the Robert C. Byrd Green Bank Telescope (GBT) in shared-risk observing time. In this paper, we show remarkable detections of abundant \hcn\, toward multiple locations within the cluster-forming region, and present initial analysis of the \amm\, emission with respect to the \hcn\, detections. In a future paper, we will present the \amm\, data in more detail. In \S \ref{sec:obs}, we describe the observations and data reduction, and present the molecular line maps and spectra in \S \ref{sec:results}. In \S \ref{sec:disc}, we show that systematic variations in the non-thermal line widths between \hcn\, and \amm, as well as the presence of coherent velocity gradients in the \hcn\, emission, suggest that the \hcn\, is largely tracing material recently accreted onto dense filaments and cores in Serpens South. In addition, we determine relative abundances of \hcn\, and \amm\, that are consistent with chemically `young' gas. We summarize our results in \S \ref{sec:summary}. 

\section{Observations}
\label{sec:obs}

\begin{figure*}
\includegraphics[width=19cm]{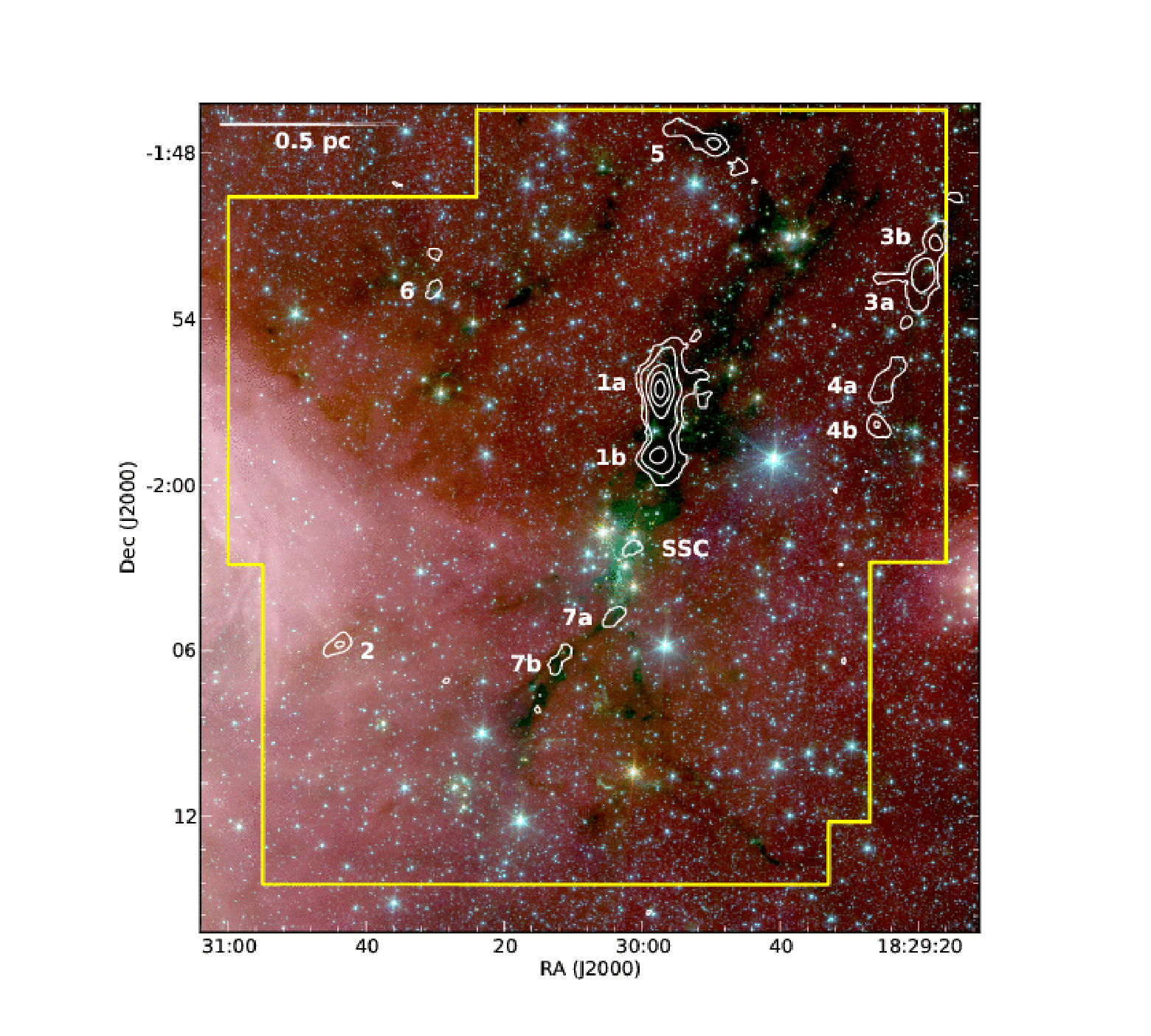}
\caption{Spitzer RGB (8\,\micron, 4.5\,\micron, 3.6\,\micron) image of the Serpens South protocluster \citep{gutermuth08}. White contours show integrated \hcn\, $J = 21-20$ emission at 0.15~K~\kms\, ($4~\sigma$), 0.3~K~\kms, 0.6~K~\kms, 0.9~K~\kms, and 1.2~K~\kms. Individual \hcn\, emission peaks discussed in the text are labeled. The physical scale assuming $d = 260$~pc is shown. Yellow contours show the map extent where the rms noise per velocity channel of width 0.15~\kms\, is $\sigma < 0.1$~K. \label{fig:hc7n_m0}}
\end{figure*}

Observations of the \amm\, $(J,K) =$ (1,1) and (2,2) inversion transitions (rest frequencies of 23.694\,GHz and 23.723\,GHz, respectively) and the \hcn\, $J = 21-20$ rotational line \citep[23.68789\,GHz;][]{kroto78} toward the Serpens South region were performed using the KFPA at the GBT between November 2010 and April 2011 in shared risk time. The total map extent is approximately 30\arcmin\, by 31\arcmin, or 2.3\,pc $\times$ 2.3\,pc assuming a distance of 260\,pc to Serpens South, with an angular resolution of $\sim 32$\arcsec\, (FWHM) or 0.04~pc. The mapped region is outlined in yellow in Figure \ref{fig:hc7n_m0}, over a three-colour (RGB) Spitzer image (blue - 3.6\,\micron; green - 4.5\,\micron; red - 8\,\micron) of the Serpens South protocluster and its surroundings \citep{gutermuth08}. The GBT spectrometer was used as the backend. We centred the first 50~MHz IF, which was observed with all seven beams, such that both the \amm\, (1,1) and (2,2) and the \hcn\, $J = 21-20$ emission lines were within the band. A second 50~MHz IF was centered on the \amm\, (3,3) line and was observed with a single beam. The \amm\, (3,3) data will be discussed in a future paper. Each IF contained 4096 channels with 12.2~kHz frequency resolution, or 0.15~\kms\, at 23.7~GHz. 

Since the telescope scan rate is limited by the rate at which data can be dumped at the GBT (where data must be dumped every 13\arcsec\, to ensure Nyquist sampling at 23\,GHz), the large map was split into ten sub-maps which were observed multiple times each. Each sub-map was made by scanning the array in rows in right ascension (R. A.), spacing subsequent scans by 13\arcsec\, in declination (Decl.). The sub-maps have similar sensitivity, although weather variations between different observing dates ensured some spread in the rms noise between sub-maps (see below). The data were taken in position-switching mode, with a common off position (R.A. 18:29:18, Decl. -2:08:00) that was checked for emission to the $T_{MB} \sim 0.1$\,K level in the \amm\, (1,1) transition. 

The data were reduced and imaged using the GBT KFPA data reduction pipeline (version 1.0) and calibrated to $T_{MB}$ units, with the additional input of relative gain factors for each of the beams and polarizations derived from standard observations (listed in Table \ref{tab:gains}). The absolute calibration accuracy is estimated to be $\sim 10$~\%. The data were then gridded to 13\arcsec\, pixels in AIPS. Baselines were fit with a second order polynomial. The mean rms noise in the off-line channels of the resulting \amm\, (1,1), (2,2), and \hcn\, data cubes is 0.06\,K per 0.15~\kms\, velocity channel, with higher values ($\sim 0.1$\,K) near the map edges where fewer beams overlapped. In general, the noise in the map is consistent, with a 1-$\sigma$ variation of 0.01\,K in the region where all the KFPA beams overlap. 

\begin{table*}
\begin{minipage}{\textwidth}
\begin{center}
\caption{KFPA Beam Gains }
\begin{tabular}{lcccccccc}
\hline
 &  & \multicolumn{7}{c}{Beams} \\
Obs Date & Polarization & 
1 & 2 & 3 & 4 & 5 & 6 & 7 \\
\hline
2010 Sep $-$ 2011 Jan\footnote{GBT Memo \#273; https://safe.nrao.edu/wiki/pub/GB/Knowledge/GBTMemos/GBTMemo273-11Feb25.pdf} & LL & 1.815 & 1.756 & 1.958 & 1.882 & 2.209 & 2.145 & 2.524 \\
                                       & RR & 2.008 & 1.842 & 1.947 & 2.00 & 2.167 &  2.145 & 2.446 \\
2011 Mar $-$ 2011 Apr\footnote{G. Langston, private communication}  & LL & 1.631 & 1.578 & 1.756 & 1.692 & 1.985 & 1.933 & 2.268 \\
				  & RR & 1.805 & 1.655 & 1.749 & 2.00 & 1.948 & 1.927 & 2.198 \\
\hline
\end{tabular}
\label{tab:gains}
\end{center}
\end{minipage}
\end{table*}

\section{Results}
\label{sec:results}

We show in Figure \ref{fig:hc7n_m0} the \hcn\, $J=21-20$ integrated intensity overlaid as contours over the Spitzer RGB image. We find significant \hcn\, emission toward multiple regions in Serpens South. Individual \hcn\, peaks are labeled 1 through 7 in order of decreasing peak line brightness temperature, with adjacent, potentially-connected sources identified as, e.g., 1a and 1b, while the emission associated with the protostellar cluster itself is labeled SSC. Contours begin at 4-$\sigma$ (0.15\,K\,\kms). We list in Table \ref{tab:sourceDetections} the R. A. and Decl. of the location of peak line brightness toward each clump identified in Figure \ref{fig:hc7n_m0}, along with the peak line brightness in $T_{MB}$. 

The strongest and most extended \hcn\, emission follows the dark 8\,\micron\, absorption feature that runs north of the central protocluster, and separates into two \hcn\, integrated intensity maxima (clumps 1a and 1b). A small \hcn\, clump is seen in the east (clump 2), with relatively bright ($\sim 1$\,K) emission and very narrow line widths ($\sim 0.18$~\kms). In the west, \hcn\, forms a filament-like feature with peak line strengths of $\sim 0.5 - 0.8$\,K (clumps 3a, 3b, 4a, and 4b). Similar line strengths are also found in the extended \hcn\, feature to the north (clump 5). Several additional \hcn\, detections are found toward the dark, narrow filament extending south-east of the central protocluster (clumps 7a and 7b), and toward some 8\,\micron\, absorption features in the north-east. While not shown, \amm\, is detected over most of the mapped area, with strong emission correlating well with the continuum, and fainter emission extending between the filaments. Small offsets between the \amm\, and continuum emission peaks are present toward some locations, as seen previously in \amm\, studies of other star-forming regions \citep{friesen09}. 

\begin{figure*}
\centering
\includegraphics*[trim=0 0 150 0, clip, height=0.5\textheight]{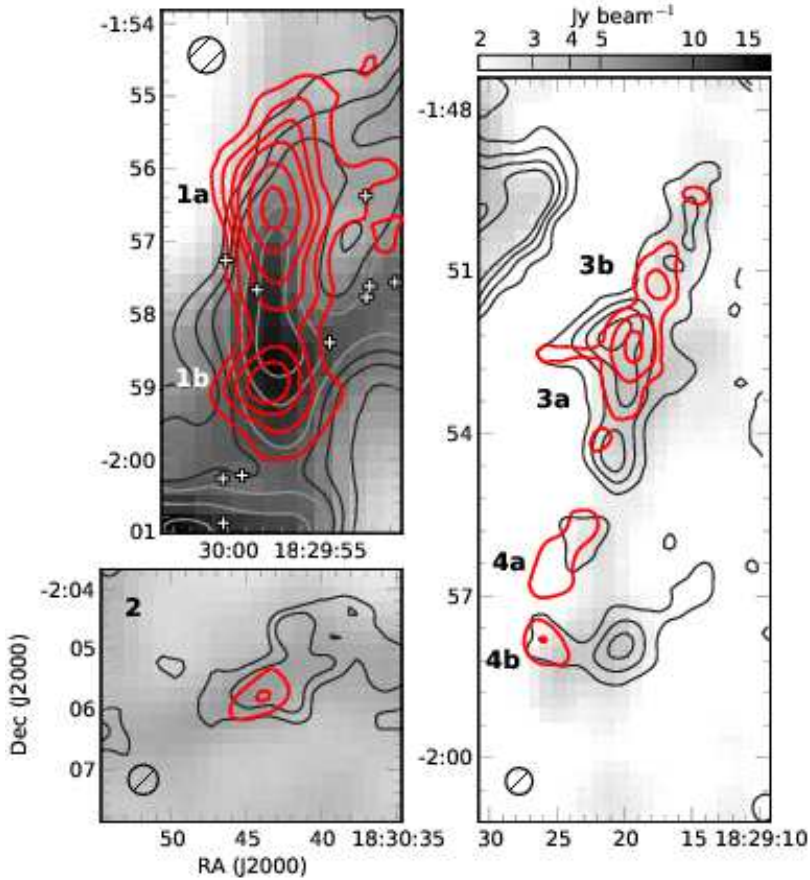}
\caption{Herschel SPIRE 500\,\micron\, dust continuum emission \citep[greyscale;][]{andre10} toward \hcn\, clumps 1a, 1b, 2, 3a, 3b, 4a, and 4b identified in Figure \ref{fig:hc7n_m0}. Overlaid are \amm\, (1,1) integrated intensity contours (dark and light grey contours) and \hcn~21-20 integrated intensity contours (red, at intervals of 0.15~K~\kms, or $4~\sigma$). The 32\arcsec\, FWHM GBT beam at 23~GHz is shown by the hashed circle in each subplot. The Herschel beam at 500~\micron\, is 36\arcsec. The Herschel data are shown to illustrate the continuum features in areas not mapped by AzTEC/ASTE at 1~mm.  White crosses show Class 0 and Class I protostar locations (Gutermuth et al., in preparation). \label{fig:hc7n_zoom1}}
\end{figure*}

\begin{figure*}
\centering
\includegraphics[width=16cm]{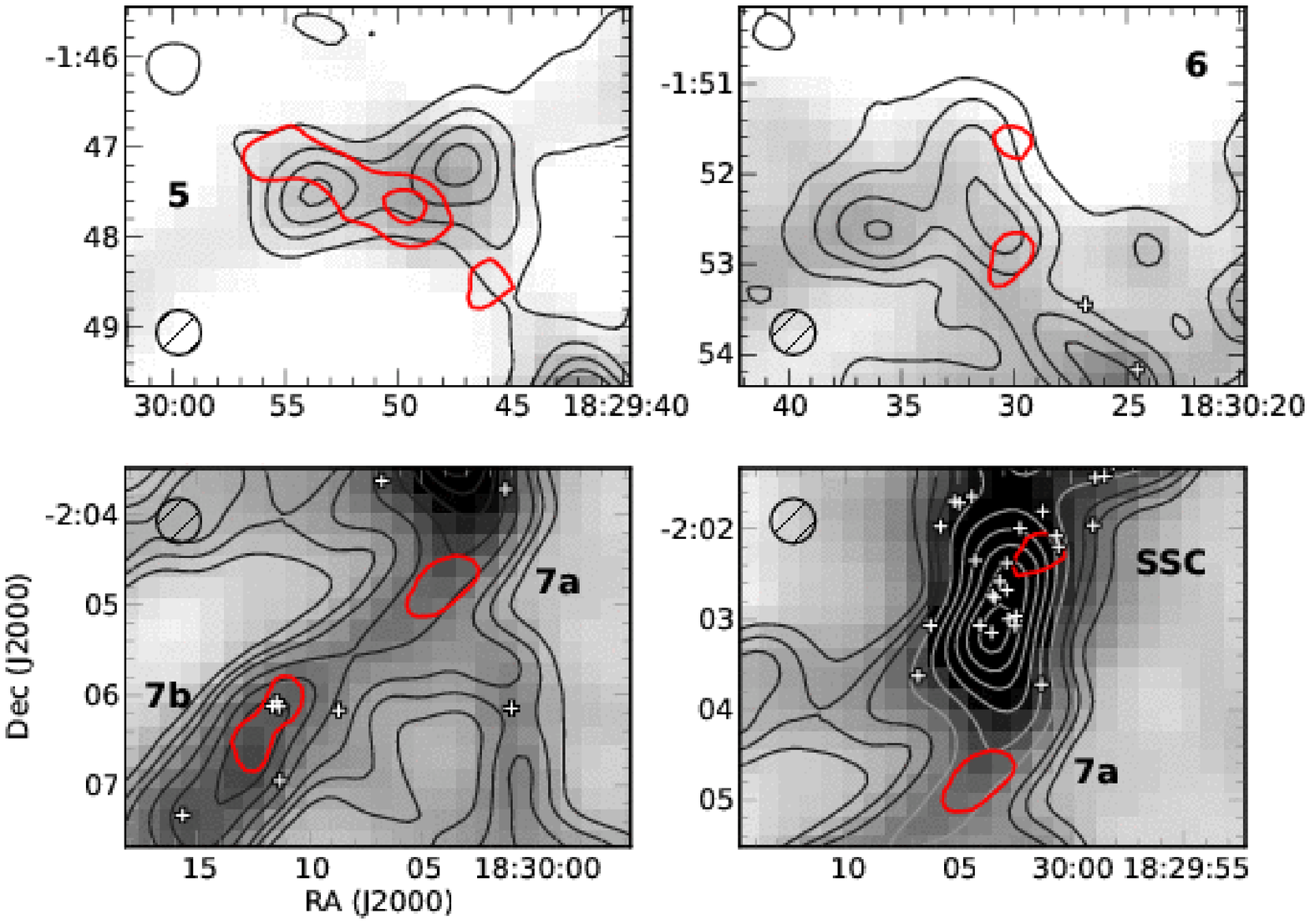}
\caption{Herschel SPIRE 500\,\micron\, dust continuum emission \citep[greyscale;][]{andre10} toward \hcn\, clumps 5, 6, 7a, 7b, and SSC identified in Figure \ref{fig:hc7n_m0}. Contours show \amm\, (1,1) integrated intensity as in Figure \ref{fig:hc7n_zoom1}. The 32\arcsec\, FWHM GBT beam at 23~GHz is shown by the hashed circle in each subplot.  White crosses show Class 0 and Class I protostar locations (Gutermuth et al., in preparation). \label{fig:hc7n_zoom2}}
\end{figure*}

Figures \ref{fig:hc7n_zoom1} and \ref{fig:hc7n_zoom2} show in greater detail the \hcn\, detections relative to both the thermal continuum emission from dust \citep[500~\micron\, emission from the Herschel SPIRE instrument at 36\arcsec\, angular resolution, taken with the Herschel Gould Belt survey;][]{andre10} and \amm\, integrated intensity (dark and light grey contours).  Herschel images of the dust continuum toward Serpens South are presented in \citet{bontemps10} and \citet{konyves10}.  We also show the locations of Spitzer-identified Class 0 and Class I protostars (Gutermuth et al., in preparation; based on the identification and classification system in \citealt{gutermuth09}). Strikingly, \hcn\, emission strongly avoids regions with numerous protostars, with the exception of a faint integrated intensity peak toward the SSC itself (labeled SSC). Few protostars are seen within the \hcn\, contours, including toward the SSC. Interestingly, \hcn\, is also not detected toward multiple dark features in Serpens South with no embedded sources. 

Figures \ref{fig:hc7n_zoom1} and \ref{fig:hc7n_zoom2} clearly show variations in the distribution of \hcn\, emission and structures traced by both \amm\, and the dust, with some spatial correspondence between all three species at low emission levels, but with emission peaks that are often distinctly offset from each other. For example, \hcn\, clumps 1a and 1b lie directly to the north and south of a filamentary ridge that is one of the strongest features in both the \amm\, and continuum maps outside of the central cluster. \hcn\, clump 2 shows the best correlation between \hcn, dust, and \amm\, emission. \hcn\, clumps 3a, 3b, 4a, and 4b all follow a faint filamentary dust structure west. The 4a and 4b peaks are offset to the east of the structure delineated by the continuum emission, while the \hcn\, and \amm\, emission contours overlap at low levels. \hcn\, clump 5 shows an elongated structure that runs between two \amm\, integrated intensity maxima, themselves within an elongated continuum structure. Two integrated intensity maxima, labeled as \hcn\, clump 6, bracket an \amm\, peak that also follows the large-scale dust emission. The 7a and 7b \hcn\, clumps lie along the filament extending to the south-east from the central cluster, with emission peaks offset from the \amm\, maxima. While few maps of \hcn\, emission have been published, observations toward TMC-1 show the \hcn\, emission peak offset by $\sim 7$\arcmin\, from the \amm\, emission peak \citep{olano88}, while maps of HC$_3$N show integrated intensity peaks offset from the continuum peak in starless cores, or the protostar location in protostellar cores \citep[e.g., L1512 and L1251A, respectively][]{cordiner11}. 

\begin{figure*}
\includegraphics{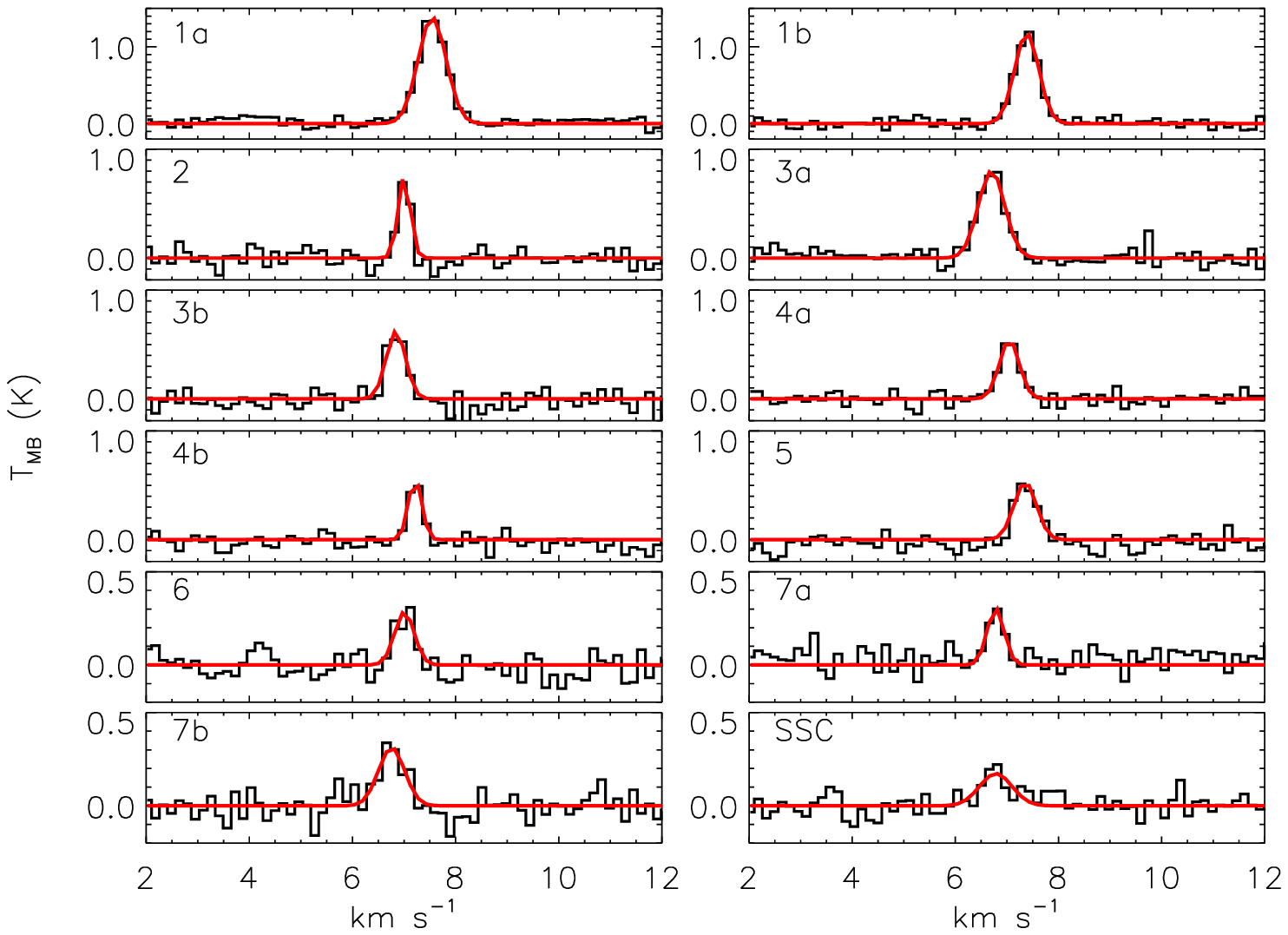}
\caption{\hcn\, 21-20 spectra (black) and Gaussian fits (red) toward the line emission peak of the \hcn\, clumps identified in Figure \ref{fig:hc7n_m0}.   \label{fig:spec}}
\end{figure*}

While many of the \hcn\, clumps are identified only by single contours of $4\sigma$ levels in integrated intensity maps, the detections are indeed significant. In Figure \ref{fig:spec}, we present the \hcn\, spectra at the locations of peak line emission toward each of the identified clumps.

\subsection{\amm\, line fitting}
\label{sec:nh3_fit}

We simultaneously fit the \amm\, (1,1) and (2,2) spectra at each pixel in the maps with a custom Gaussian hyperfine structure fitting routine written in IDL. The routine used was modified from that described in detail in \citet{friesen09}, where the \amm\, (1,1) line was fit using the full hyperfine analysis, while the \amm\, (2,2) line was fit separately by a single Gaussian. Here, we create model \amm\, (1,1) and (2,2) spectra given input kinetic gas temperature, $T_K$, the line-of-sight (LOS) velocity relative to the local standard of rest (LSR), \vlsr\,, the line full-width at half maximum, $\Delta v$, the \amm\, (1,1) opacity, $\tau_{(1,1)}$, and the excitation temperature, $T_{ex}$. We then perform a chi-square minimization against the observed \amm\, emission. This fitting routine is similar to that detailed by \citet{rosolowsky08}. The model assumes equal excitation temperatures for the \amm\, (1,1) and (2,2) lines, and for each hyperfine component. The model spectra are created assuming the \amm\, (1,1) and (2,2) lines also have equal LOS velocities and equal Gaussian line widths. The opacity of the (2,2) line is calculated relative to the (1,1) line following \citet{hotownes}. The expected emission frequencies and emitted line fractions for the hyperfine components were taken from \citet{kukolich}. We additionally assume the line emission fills the beam (the best-fit $T_{ex}$ will be a lower limit if the observed emission does not entirely fill the beam). 

In our analysis, we mask the fit results where the signal-to-noise ratio, $S/N < 4$ in the \amm\, (1,1) line for \vlsr\, and $\Delta v$, and mask where $S/N < 5$ in the \amm\, (2,2) line for $T_K$ (and consequently results that depend on $T_K$). We find that this cut ensures that both the \amm\, (1,1) and (2,2) lines are  detected with enough S/N that the kinetic gas temperatures are determined through the hyperfine fits to a precision of $\sim 1-2$\,K. Where two distinct \amm\, components are found along the line of sight, the routine fits the \amm\, hyperfine structure corresponding to the strongest component. In pixels where the line strengths of both components are nearly equal, we thus find larger returned line widths. This effect is only seen, however, toward small areas in the map, and is not an issue for any regions where \hcn\, is detected. The \amm\, column density is derived from the returned parameters following \citet{friesen09}. 

In the following discussion, we focus only on the \amm\, kinematics, column density and temperature as these properties relate to the detected \hcn\, emission. A detailed study of the dense gas structure and kinematics in Serpens South as traced by \amm\, will be presented in an upcoming paper. 

\begin{table}
\caption{HC$_7$N Emission Peak Locations }
\begin{tabular}{lccc}
\hline
Clump ID & R.A. & Decl. & $T_{peak, MB}$ \\
 & J2000 & J2000 & K \\
\hline
 1a & 18:29:57.9 & -1:56:19.0 &  2.11 \\
 1b & 18:29:57.0 & -1:58:55.0 &  1.47 \\
  2 & 18:30:43.8 & -2:05:51.0 &  1.13 \\ 
 3a & 18:29:19.7 & -1:52:25.0 &  0.82 \\ 
 3b & 18:29:18.0 & -1:51:20.0 &  0.72 \\ 
 4a & 18:29:24.9 & -1:56:32.0 &  0.74 \\ 
 4b & 18:29:25.8 & -1:58:03.0 &  0.72 \\ 
  5 & 18:29:50.9 & -1:47:39.0 &  0.66 \\ 
  6 & 18:30:29.9 & -1:53:04.0 &  0.40 \\ 
  7a & 18:30:03.9 & -2:04:46.0 &  0.36 \\
 7b & 18:30:11.7 & -2:06:17.0 &  0.33 \\
SSC & 18:30:01.3 & -2:02:10.0 &  0.23 \\
\hline
\end{tabular}
\label{tab:sourceDetections}
\end{table}

\subsection{\hcn\, line fitting}
\label{sec:hcn_fit}

The rotational lines of \hcn\, exhibit quadrupole hyperfine splitting, with line spacing approximately $e q Q/4J^2$, where $eqQ = 4.29$~MHz for \hcn\, \citep{mccarthy00}. The resulting splitting for the \hcn\, $J=21-20$ transition is $\sim 2.5$~kHz, and is unresolved by our spectral resolution of 12~kHz. We therefore fit the \hcn\, emission with a single Gaussian to determine \vlsr\,, $\Delta v$, and the peak amplitude of the emission where detected with a signal-to-noise $S/N \geq 4$ across the map. Figure \ref{fig:spec} shows the \hcn\, $J = 21-20$ spectra toward the peak emission locations of the integrated intensity maxima identified in Figure \ref{fig:hc7n_m0} with overlaid Gaussian fits. Even toward the faint detections in the integrated intensity map, the lines are well-fit by single Gaussians. We list in Table \ref{tab:clumpValues} the mean and $1\sigma$ variation in \vlsr\,, line width FWHM $\Delta v$, and peak brightness temperature, $T_{MB}$, for each \hcn\, clump. 

Toward most of the \hcn\, features, we find line widths $\Delta v \sim 0.4 - 0.5$~\kms, which are similar to those found toward several other regions. For example, \hcn\, line widths toward L492 and TMC-1 are 0.56~\kms\, and 0.43~\kms, respectively, although at poorer angular resolution than in this study (1\arcmin.2 and 1\arcmin.6 versus 32\arcsec). The \hcn\, line in TMC-1 was further shown to be a composite of several velocity components with more narrow lines \citep{dickens01}. The protostellar source L1527 shows similar line widths to those we find in Serpens South, at the same angular resolution \citep{sakai08}. Many previous \hcn\, detections toward low mass starless and protostellar sources, however, show significantly smaller line widths, with $\Delta v \sim 0.2 - 0.3$~\kms\, \citep[e.g.,][]{cernicharo86}.  

We determine the column density from the integrated intensity, $W$, assuming the line is optically thin and in local thermodynamic equilibrium following \citet{olano88}: 

\begin{equation}
N = \frac{3 k^2}{8 \pi^3 h B} \frac{1}{\mu^2 (J_l+1) \nu} \frac{T_{ex} \, J_\nu(T_{ex}) \exp(E_u/ (k T_{ex}) )}{J_\nu(T_{ex}) - J_\nu(T_{bg})} W
\end{equation}

Here, $B = 564.0$\,MHz is the rotational constant \citep{mccarthy00}, $\mu = 4.82$~Debye is the dipole moment, $J_l$ is the lower rotational level, $\nu$ is the transition frequency, and $E_u/k = 11.37$\,K is the energy of the upper level of the transition. $J_\nu (T_K)$ and $J_\nu (T_{bg})$ are the Rayleigh-Jeans equivalent excitation and background temperatures ($T_{bg} = 2.73$\,K), where $J_\nu (T) = h \nu / k \, \times \, (\exp(h\nu/kT) -1)^{-1}$. We set the excitation temperature equal to the kinetic gas temperature derived from the \amm\, (1,1) and (2,2) hyperfine line fit analysis. We calculate the formal uncertainty in $W$, $\sigma_W = \Delta v_{res} (N_{ch})^{1/2} rms$, where $\Delta v_{res}$ is the spectral resolution in km\,s$^{-1}$, $N_{ch}$ is the number of channels in the integrated area, and $rms$ is the rms noise level. Given the noise level in the map and using a typical temperature $T_K = 11$\,K (the mean kinetic gas temperature traced by \amm\, where \hcn\, is detected is 10.8\,K; see \S \ref{sec:disc}), the 3-$\sigma$ lower limit in \hcn\, column density is $\sim 2 \times 10^{12}$\,cm$^{-2}$. The $\sigma$ values used to calculate this limit do not include the $\sim 10$\,\% overall calibration uncertainty, or the uncertainty introduced by setting $T_{ex} = T_K$. We expect $T_{ex} = T_K$ is a good assumption, however, as \hcn\, should thermalize rapidly at densities $n > 10^3$\,\cc\, due to its large collisional cross section (see \S \ref{sec:chem}). A decrease in $T_{ex}$ of 2\,K from the mean $T_K$ value would increase the resulting column density by only 5\,\%. Alternatively, \hcn\, may be excited in gas that is warmer than that traced by \amm. Increasing the excitation temperature by $2-4$~K, however, results in a decrease in $N(\mbox{\hcn})$ of only $3-4$~\%. 

Where detected, we find typical \hcn\, column densities $N(\mbox{\hcn}) \sim 3-6 \times 10^{12}$\,cm$^{-2}$ outside of the main north-south filament, where $N(\mbox{\hcn}) \sim 1-3 \times 10^{13}$\,cm$^{-2}$ in clumps 1a and 1b. The larger column density peaks in 1a and 1b are similar to those found in TMC-1, where large cyanopolyynes were first detected in star-forming regions \citep{kroto78}. The lower range of column densities is similar to values found in low-mass starless and protostellar cores where \hcn\, has been detected previously. For example, \citet{cordiner11} find $N(\mbox{\hcn}) = 1.9 \pm 0.1 \times 10^{12}$\,cm$^{-2}$ toward the starless core L1512, and $N(\mbox{\hcn}) = 4.7 \pm 0.4 \times 10^{12}$\,cm$^{-2}$ toward a location 40\arcsec\, offset from the L1251A IRS3 Class 0 protostar. 

\begin{table*}
\begin{minipage}{\textwidth}
\begin{center}
\caption{HC$_7$N Clump Properties}
\begin{tabular}{lccccccc}
\hline
Clump ID & $v_{LSR}$ & 
$\Delta v$ & $T_{MB}$ & 
$\sigma_{NT}$\footnote{Non-thermal line widths are derived assuming a gas temperature equal to the kinetic temperature derived from \amm\, hfs fitting.} & 
$N(\mbox{\hcn})$ & 
$X(\mbox{\hcn})$\footnote{\hcn\, abundances are derived as $N(\mbox{\hcn})/N(\mbox{H}_2)$, where $N(\mbox{H}_2)$ is determined from 1.1~mm continuum emission (Gutermuth et al. in prep). Several \hcn\, clumps are outside the boundaries of the millimeter continuum map.} &
$[\mbox{\amm}]\,/\,[\mbox{\hcn}]$ \\
 & \kms & \kms & K & \kms & 
$\times 10^{12}$\,cm$^{-2}$ & 
$\times 10^{-10}$\,cm$^{-2}$ &  \\
\hline
   1a &  7.36 (0.25) &  0.57 (0.15) & 0.64 (0.48) & 0.24 (0.07) &  8.9 (6.4) &  3.2 (2.6) &   176 ( 107) \\
   1b &  7.32 (0.11) &  0.58 (0.13) & 0.53 (0.36) & 0.25 (0.05) &  6.6 (3.6) &  0.9 (0.4) &   283 ( 111) \\
    2 &  7.01 (0.02) &  0.18 (0.06) & 0.59 (0.25) & 0.08 (0.02) &  4.6 (1.5) &  3.1 (0.8) &    69 (  41) \\
   3a &  6.66 (0.20) &  0.55 (0.14) & 0.47 (0.18) & 0.23 (0.06) &  5.6 (2.2) &  \nodata &    77 (  32) \\
   3b &  6.86 (0.07) &  0.52 (0.07) & 0.47 (0.14) & 0.22 (0.03) &  4.4 (1.5) &  \nodata &   159 ( 387) \\
   4a &  7.05 (0.08) &  0.43 (0.07) & 0.39 (0.15) & 0.18 (0.03) &  3.9 (1.0) &  7.4 (3.1) &    75 (  60) \\
   4b &  7.22 (0.04) &  0.35 (0.07) & 0.41 (0.14) & 0.15 (0.03) &  3.4 (1.1) &  4.0 (3.2) &    79 (  59) \\
    5 &  7.40 (0.17) &  0.57 (0.13) & 0.36 (0.13) & 0.24 (0.06) &  3.9 (1.2) &  \nodata &   152 (  65) \\
    6 &  7.04 (0.06) &  0.46 (0.19) & 0.29 (0.05) & 0.19 (0.08) &  4.3 (1.2) &  3.5 (3.2) &   207 (  71) \\
   7a &  6.76 (0.10) &  0.52 (0.38) & 0.26 (0.07) & 0.22 (0.16) &  4.5 (0.7) &  1.4 (0.3) &   283 (  37) \\
   7b &  6.87 (0.59) &  0.61 (0.19) & 0.27 (0.04) & 0.26 (0.08) &  4.0 (1.1) &  0.8 (0.3) &   736 ( 201) \\
  SSC &  6.84 (0.11) &  0.63 (0.30) & 0.18 (0.03) & 0.27 (0.13) &  3.1 (0.4) &  0.5 (0.1) &   557 ( 130) \\
\hline
\end{tabular}
\label{tab:clumpValues}
\end{center}
\end{minipage}
\end{table*}

\subsection{Abundances}
\label{sec:abundances}

We calculate the abundance along the line-of-sight of \amm\, and \hcn\, relative to H$_2$ by determining the H$_2$ column density from 1.1\,mm thermal continuum data observed toward Serpens South with the 144-pixel bolometer camera AzTEC on the Atacama Submillimeter Telescope Experiment (ASTE; Gutermuth et al., in prep). The AzTEC data, which have also been presented in \citet{nakamura11}, \citet{sugitani11}, and \citet{kirk13}, are well-matched in angular resolution with the GBT data (28\arcsec\, FWHM vs. 32\arcsec\, FWHM). The AzTEC map covers a slightly smaller region than the GBT observations, so we are not able to determine the H$_2$ column density toward the \hcn\, clumps 3a, 3b, and 6.  We derive $N(\mbox{H}_2)$ following the standard modified blackbody analysis, $N(\mbox{H}_2) = S_\nu /(\Omega_m \mu m_{\mbox{\scriptsize{H}}} \kappa_\nu B_\nu(T_d))$, where $S_\nu$ is the continuum flux density, $\Omega_m$ is the beam solid angle, $\mu = 2.33$ is the mean molecular weight, $m_{\mbox{\scriptsize{H}}}$ is the mass of hydrogen, $\kappa_\nu$ is the dust opacity per unit mass, and $B_\nu (T_d)$ is the Planck function at the dust temperature $T_d$. We set the dust opacity at 1.1\,mm, $\kappa_{1.1{\mbox{\scriptsize{mm}}}} = 0.0114$~cm$^{2}$~g$^{-1}$, interpolated from the \citet{ossen94} dust model for grains with coagulated ice mantles \citep{schnee09}. We set the dust temperature equal to the kinetic gas temperature as determined by \amm. In general, we expect this second assumption to introduce little error, since the dust and gas are likely coupled at the densities typical of the Serpens South filaments \citep{goldsmith01}. 

\begin{figure*}
\includegraphics[width=0.45\textwidth]{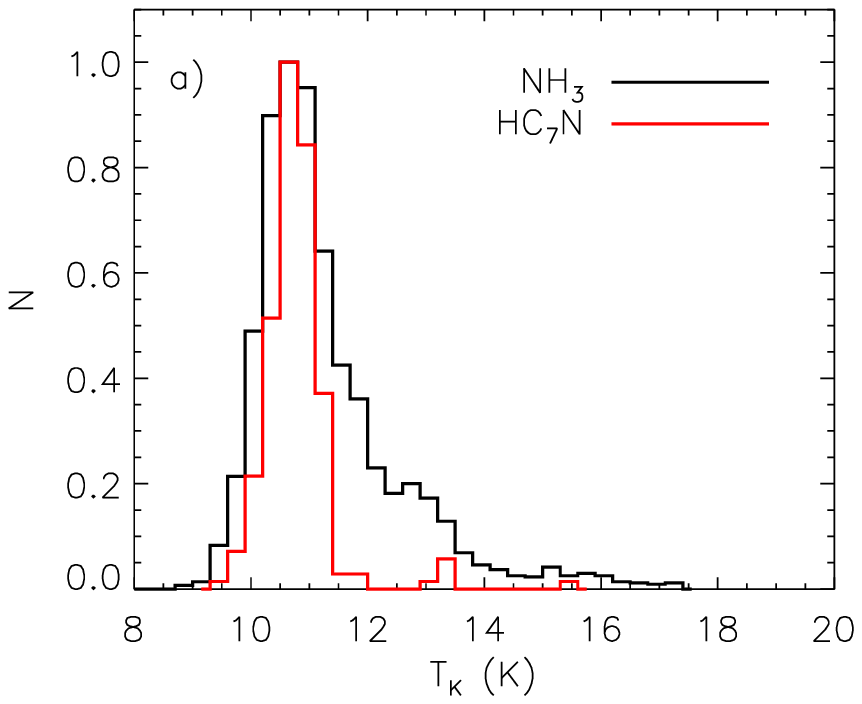}
\includegraphics[width=0.45\textwidth]{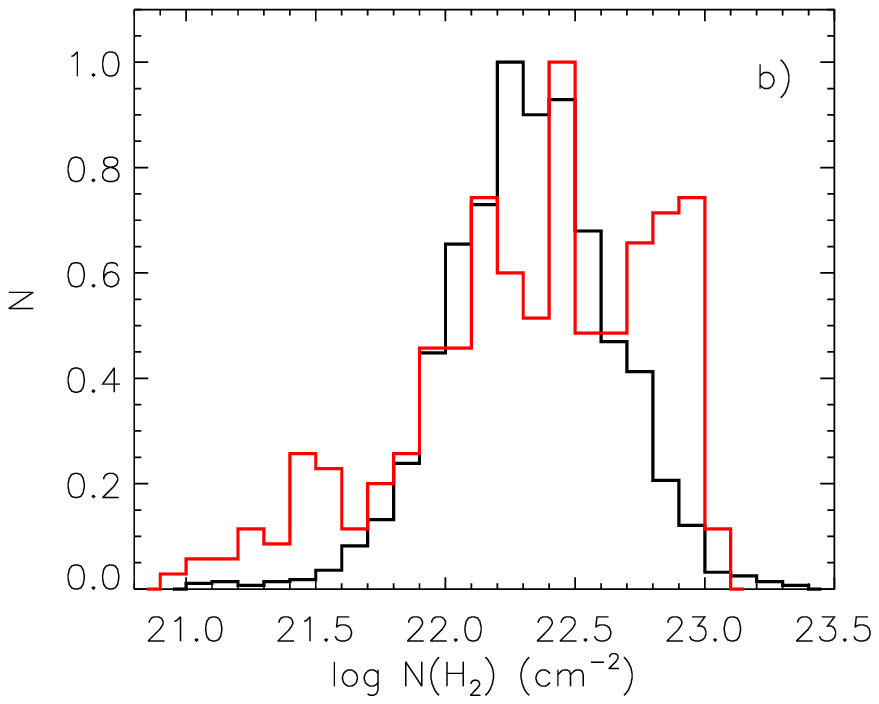}
\caption{a) Normalized histogram showing the gas temperature distribution where \hcn\, (red) and \amm\, (black) are detected in Serpens South. Although both distributions peak at $T_K \sim 10.8$\,K, \hcn\, is almost entirely found at low temperatures (mean $T_K = 10.8$\,K, $\sigma_{T} = 0.6$\,K), with the exception of a few warmer points associated with the SSC \hcn\, clump. In contrast, \amm\, also traces warmer material (mean $T_K = 11.3$\,K, $\sigma_T = 1.3$\,K). b) Normalized histogram showing the H$_2$ column density distribution where \hcn\, (red) and \amm\, (black) are detected in Serpens South. Note in both panels that some \hcn\, emission contours extend beyond the regions where $T_K$ or $N(\mbox{H}_2)$ are fit well. In a), these points are most likely at low $T_K$; in b), these points are most likely at low $N(\mbox{H}_2)$.  \label{fig:tkHist}}
\end{figure*}

The distribution and locations of peak \hcn\, emission are frequently offset from peaks in the \amm\, emission, and also from peaks in the (sub)millimeter continuum emission. Table \ref{tab:clumpValues} gives the mean and standard deviation of the $[\mbox{\amm}]\,/\,[\mbox{\hcn}]$ ratio for each \hcn\, clump. Several clumps show mean $[\mbox{\amm}]\,/\,[\mbox{\hcn}] \sim 70 - 90$, specifically \hcn\, clump 2, as well as clumps 3a, 3b, 4a, and 4b, which lie together to the west of the main SSC filament. The rest have mean $[\mbox{\amm}]\,/\,[\mbox{\hcn}] \sim 200-300$, apart from the 7b clump, which has the highest detected \amm\, abundance relative to \hcn\, (mean $[\mbox{\amm}]\,/\,[\mbox{\hcn}] = 741$). In most of the \hcn\, clumps, the relative abundance of \amm\, to \hcn\, is similar to or lower than that seen toward other star-forming regions with \hcn\, and \amm\, detections. In other regions, \citet{cernicharo86} find $[\mbox{\amm}]\,/\,[\mbox{\hcn}] = 132$ toward L1495, while the column densities reported by \citet[][; collected from results of largely single-point observations by \citet{cernicharo86,hirota04,suzuki92}]{hirota06} give $[\mbox{\amm}]\,/\,[\mbox{\hcn}]$ values of 106, 52, and 14 toward L492, L1521B, and TMC-1, respectively. As discussed earlier, the peak locations of \hcn\, and \amm\, are offset toward TMC-1, and \citet{olano88} find a range of $[\mbox{\amm}]\,/\,[\mbox{\hcn}]$ values between 26 and $\gtrsim 200$ using maps of both species.

\section{Discussion}
\label{sec:disc}

\subsection{\hcn\, and \amm\, chemistry}
\label{sec:chem}

The \hcn\, $J=21-20$ and \amm\, $(J,K) = (1,1)$ inversion transitions are excited in similar conditions. The critical density of the \hcn\, transition, where collisional de-excitation rates are equal to radiative de-excitation rates, is given by $n_{cr} = A_{ul}/(v\,\sigma)$. Here, $A_{ul}$ is the spontaneous emission coefficient, $\sigma$ is the collisional cross section, and $v$ is the average velocity of the collision partners (H$_2$ and He). For \hcn\, $J = 21-20$, $A_{ul} \sim 1.6 \times 10^{-6}$~s$^{-1}$ \citep{olano88}. The collisional cross section of \hcn\, is expected to be large, and has been estimated by scaling the cross-section for HC$_3$N by the ratio of the \hcn/HC$_3$N linear lengths, giving $\sigma = 1.6 \times 10^{-14}$~cm$^2$ with an uncertainty of $\sim 25$~\% \citep{bujarr81}. Setting $v = \sqrt{kT/m}$, where $T=11$~K and $m = 2.33$~\mh, we find the resulting \hcn\, critical density $n_{cr} \sim 5 \times 10^3$~\cc. For \amm\, (1,1), $n_{cr} = 3.9 \times 10^3$~\cc\, \citep{maret09}. Given the similarity of the critical densities and the rest frequencies of the \hcn\, and \amm\, transitions, the excitation curves of the two transitions as a function of gas density, $T_{ex}$ vs. $n$, also agree well. 

In most regions in Serpens South, we find that the \hcn\, emission peaks are offset from peaks in the \amm\, integrated intensity and millimeter continuum emission (see Figures \ref{fig:hc7n_zoom1} and \ref{fig:hc7n_zoom2}). Furthermore, Figure \ref{fig:tkHist} shows that \amm\, and \hcn\, show different distributions with gas temperature (a) and with H$_2$ column density (b). \hcn\, is preferentially present toward cold locations in Serpens South, whereas \amm\, is also excited in warmer regions. In addition, \hcn\, emission is not detected over a smooth distribution of H$_2$ column density, in contrast to \amm. Instead, the \hcn\, detections are relatively overdense at both low and high column density (although the distribution is not bimodal). Differences in the observed distribution of the molecular line emission must therefore be explained through chemistry or dynamics in the cloud. In this sub-section, we discuss the chemical formation and destruction pathways, and relevant timescales, for \amm\, and \hcn. 

\begin{figure}
\includegraphics[width=0.5\textwidth]{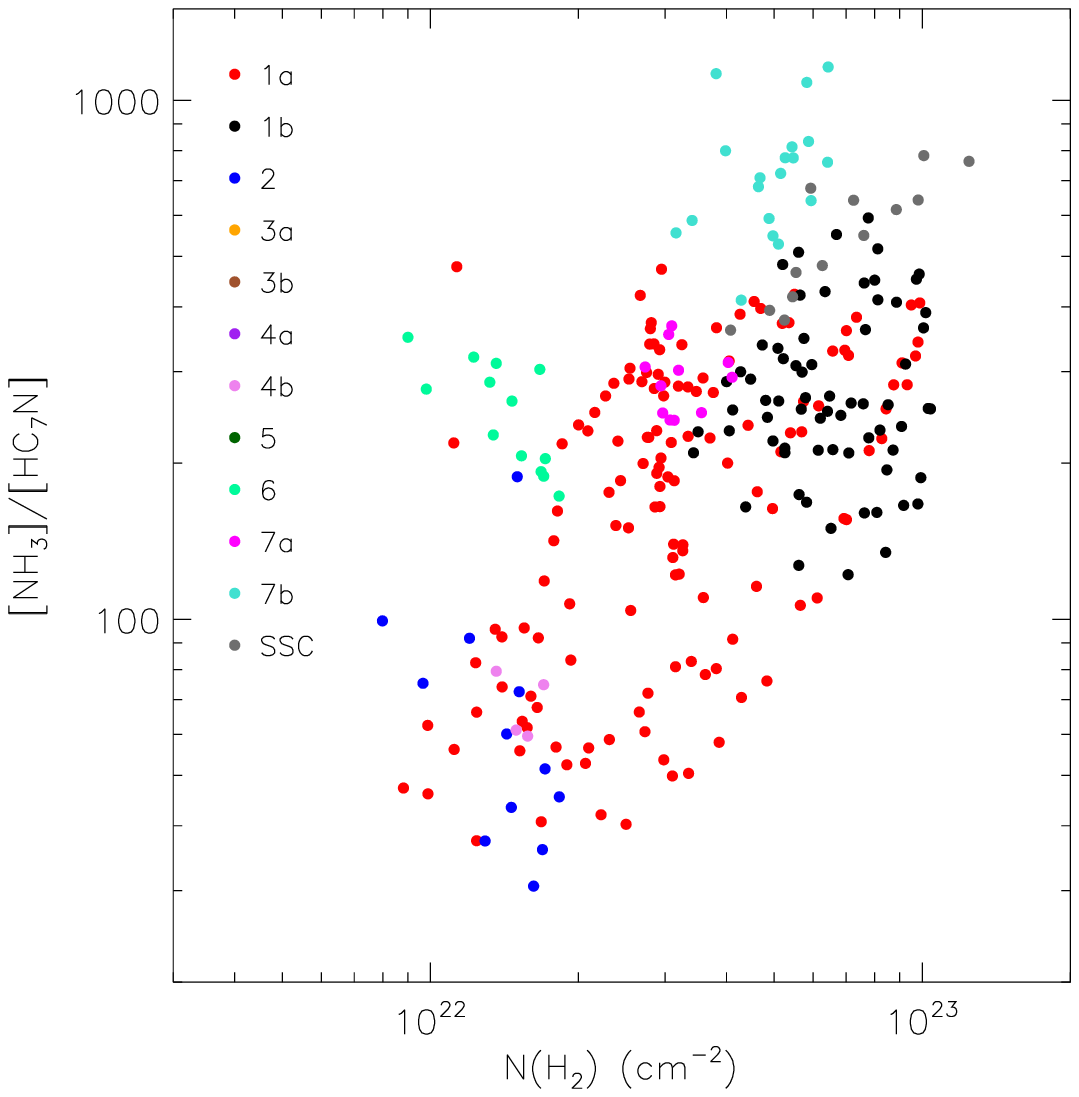}
\caption{Ratio of \amm\, to \hcn\, column density as a function of $N(\mbox{H}_2)$. Points show 13\arcsec\, pixels and are colored to show individual \hcn\, clumps. Some \hcn\, clumps (3a, 3b, 6) are outside of the 1.1~mm map and do not have $N(\mbox{H}_2)$ measurements.  The abundance ratio of \amm\, relative to \hcn\, varies by nearly two orders of magnitude over more than one order of magnitude range in $N(\mbox{H}_2)$. The \hcn\, 1a and 1b clumps, in particular, span most of the observed values, while other clumps are more localized in $[\mbox{\amm}]/[\mbox{\hcn}]$ - $N(\mbox{H}_2)$ space.  \label{fig:hvsn_column}}
\end{figure}

Since the \hcn\, emission is likely optically thin, in general the locations of greatest \hcn\, abundance are also likely offset from the \amm\, and continuum peaks. We show in Figure \ref{fig:hvsn_column} the distribution of $[\mbox{\amm}]~/~[\mbox{\hcn}]$ relative to $N(\mbox{H}_2)$, colored by \hcn\, clump (omitting those clumps outside of the millimeter continuum map). We find that the \amm\, to \hcn\, abundance increases with $N(\mbox{H}_2)$, although the scatter in this trend is large. The span in relative \amm\,/\,\hcn\, abundance is a factor of $\sim 35$, while the span in H$_2$ column density is a factor of $\sim 12$. In particular, \hcn\, clump 1a spans most of the values observed in both $[\mbox{\amm}]~/~[\mbox{\hcn}]$ and $N(\mbox{H}_2)$, while the other \hcn\, clumps show less variation in both values.  \hcn\, clump 4, which lies within a narrow, low column density filament seen in continuum emission, has both the lowest H$_2$ column density (where we are able to measure) and lowest mean ratio of $[\mbox{\amm}]/[\mbox{\hcn}]$. 

\begin{figure*}
\includegraphics{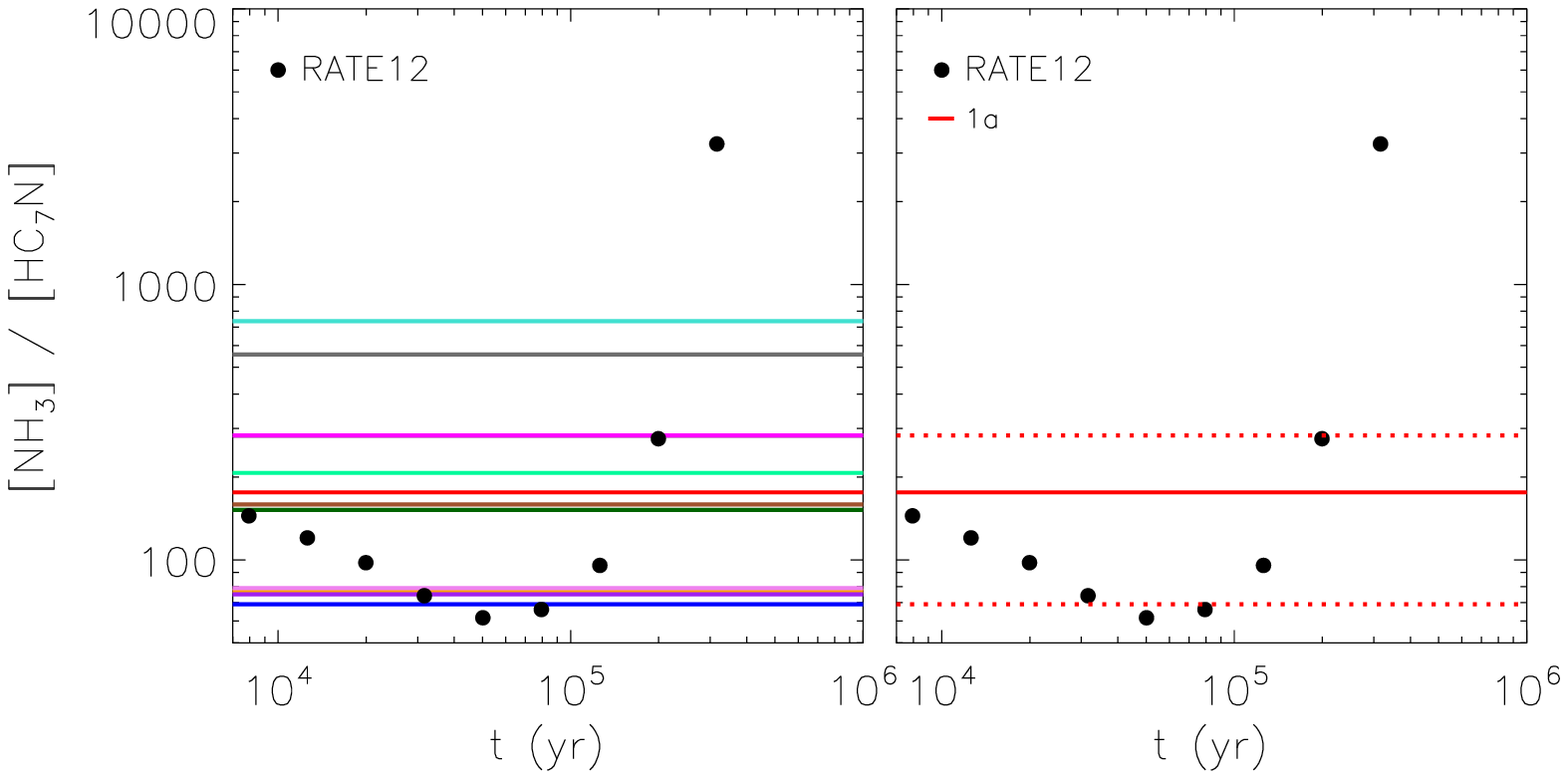}
\caption{Left: Black points show predictions from the `dark cloud' chemical model ($n = 10^4$\,\cc, $T = 10$\,K, $A_v = 10$) by \citet[][RATE12]{mcelroy13} for the $[\mbox{\amm}]/[\mbox{\hcn}]$ ratio as a function of time. Coloured lines show the mean values for individual \hcn\, clumps in Serpens South, with colors matched to those in Figure \ref{fig:hvsn_column}. Right: Same points as at left, with the red line showing the mean value for \hcn\, clump 1a only. Dashed red lines show the standard deviation around the mean.  In this region, the minimum ratio $[\mbox{\amm}]/[\mbox{\hcn}] \sim 30$ (see Figure \ref{fig:hc7nGrad}). \label{fig:r12model}}
\end{figure*}

The distribution of \hcn\, and \amm\, in Serpens South can be explained through simple chemical models of cold, dense gas, assuming that regions of higher H$_2$ column density also track greater gas volume densities. Since the formation of \amm\, is limited by neutral-neutral reactions, it has a long formation timescale at a given gas density. In contrast, chemical models show that long carbon-chain molecules like \hcn\, are produced rapidly in cold, dark clouds through ion-neutral and neutral-neutral reactions, but high abundances of these species are only found well before the models reach a steady-state chemical equilibrium \citep{herbst89}. At densities $n \sim 10^4$~\cc\, and temperatures $T \sim 10$~K, \hcn\, peaks in abundance, and then rapidly becomes less abundant due to chemical reactions after $t \sim 10^5$\,yr \citep{cwalsh09}. The trend in the relative abundances of \amm\, and \hcn\, is shown in Figure \ref{fig:r12model}, where we plot the results of the simple `dark cloud' chemical model ($T = 10$~K, $n = 10^4$~\cc) of \citet{mcelroy13} for $[\mbox{\amm}]/[\mbox{\hcn}]$ as a function of time (black points). As \hcn\, forms more quickly at early times, the $[\mbox{\amm}]/[\mbox{\hcn}]$ abundance ratio reaches a minimum at $t < 10^5$~yr, but rapidly increases at $t > 10^5$~yr as \amm\, continues to grow in abundance while \hcn\, is depleted. Overlaid on the Figure are the mean $[\mbox{\amm}]/[\mbox{\hcn}]$ values for each \hcn\, clump, which cluster toward values consistent with $t \lesssim 2 \times 10^5$~yr. 

Some models suggest that  a second cyanopolyyne abundance peak may occur at later times in the evolution of a star-forming cloud, as elements start to deplete from the gas phase through 'freezing out' onto dust grains. Specifically, the depletion of gas-phase O, which reacts destructively with carbon-chain molecules, is the main cause of this second, 'freeze out' peak \citep{brown90}. In Serpens South, however, the observed distribution of \hcn\, relative to the dust continuum and \amm\, emission, along with the kinematics discussed above, suggests that the \hcn\, is tracing material that has only recently reached sufficient density for the species to form. 

One possible exception to this interpretation is the \hcn\, SSC clump, which is co-located with the central cluster, although offset to the north-west from both the continuum and \amm\, emission peaks associated with the cluster (see Figure \ref{fig:hc7n_zoom2}). Given the large surface density of young stars in the cluster core, we may expect the core chemistry to be similar to that found in 'hot corinos', the low-mass analog of hot cores around high mass stars \citep{ceccarelli05}. In these regions, many complex organic species are found, and chemical models suggest long-chain cyanopolyynes like HC$_5$N and \hcn\, may again become abundant in the gas phase for short times \citep{chapman09}. While the kinetic gas temperature traced by \amm\, is warmer than in the filaments, it is not sufficient ($T_K \sim 15.5$~K toward the \hcn\, detection) to remove substantial amounts of molecular species from dust grain mantles as expected in a hot molecular core. The \amm\, analysis gives a mean temperature along the line-of-sight, however, and gas temperatures may indeed be greater in the cluster interior. Alternatively, comparison of the location of the \hcn\, emission relative to the protostars identified in the SSC shows that even in this cluster, only a few YSOs have been identified near the edges of the \hcn\, SSC integrated intensity contour, with none toward the \hcn\, emission peak (Gutermuth et al., in prep). This suggests that even at the cluster centre, \hcn\, is only abundant where stars have yet to form, and may again highlight chemically younger material in the central region as in the rest of the region. 

\subsection{Kinematics of \amm\, and \hcn}
\label{sec:kin}

\begin{figure*}
\includegraphics{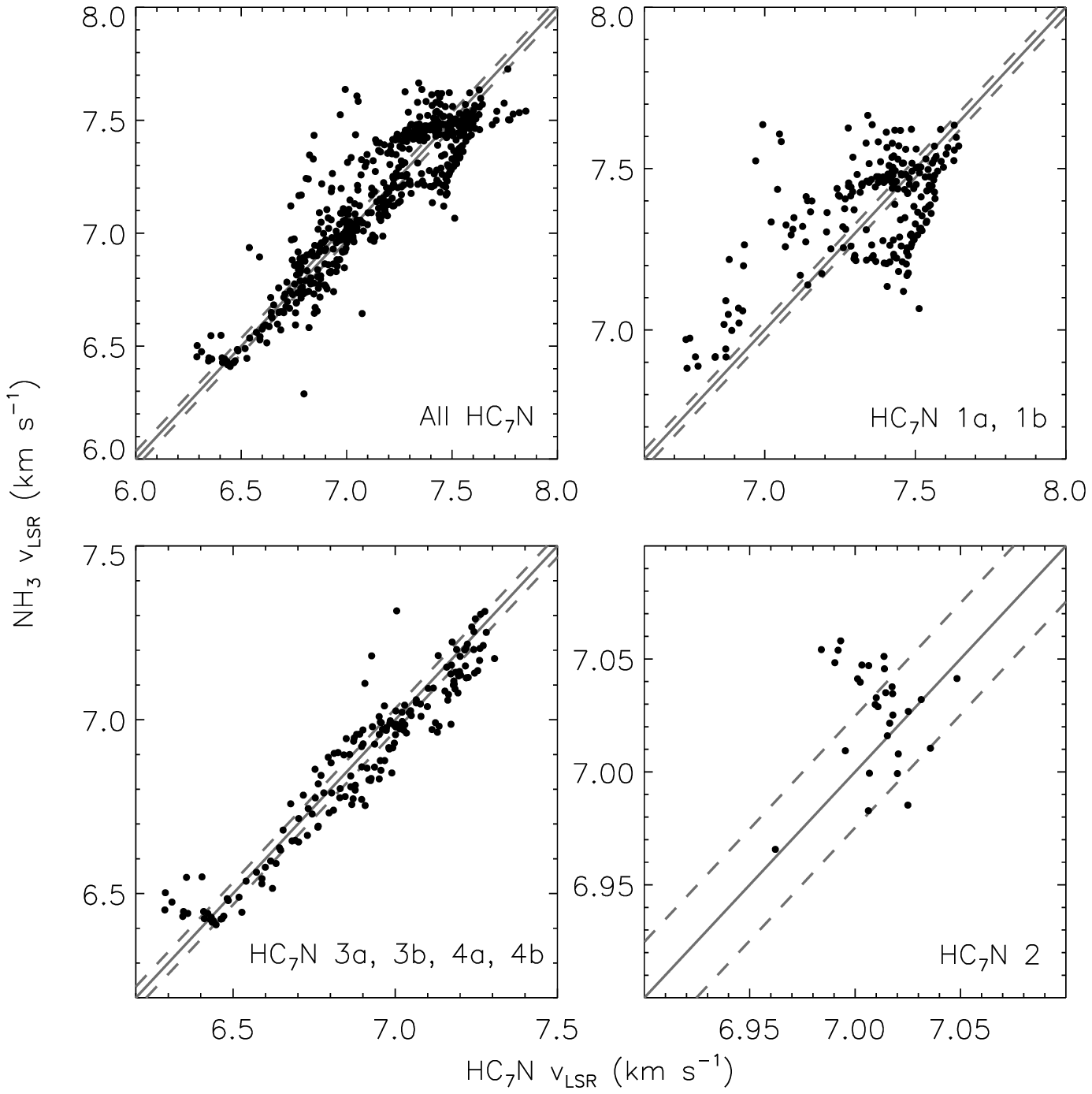}
\caption{Top left: Comparison of \amm\, and \hcn\, line centroids (\vlsr\,) for all pixels detected in both \amm\, and \hcn\, above the 4-$\sigma$ level. Points show 13\arcsec\, pixels. The solid grey line shows equality, while the dashed lines show the 1-1 relationship, plus or minus the mean uncertainty in the line centroid from the Gaussian fits to the \hcn\, lines for the data points plotted. \amm\, line centroids are determined to up to a factor of $\sim 10$ better accuracy due to the simultaneous fitting of 18 hyperfine components in the \amm\, (1,1) line. While the line centroids agree on average, significant deviations are seen toward multiple regions. Top right: As in top left but limited to \hcn\, clumps 1a and 1b as labeled in Figure \ref{fig:hc7n_m0}. Bottom left: As in top left but limited to \hcn\, clumps 3a, 3b, 4a, and 4b. Bottom right: As in top left but limited to \hcn\, clump 2. \label{fig:hcn_velo}}
\end{figure*}

\subsubsection{Differences in line centroids and non-thermal line widths}

In general, \amm\, and \hcn\, show similar LOS velocities where detected along the same line-of-sight, as shown in Figure \ref{fig:hcn_velo}, where a comparison of \amm\, and \hcn\, line centroids largely falls along the line of unity (shown as a solid grey line in the Figure). As a guide, the dashed lines show the 1-1 relationship, plus or minus the mean uncertainty in the line centroid from the Gaussian fits to the \hcn\, lines for the data points plotted. \amm\, line centroids are determined to up to a factor of $\sim 10$ better accuracy due to the simultaneous fitting of 18 hyperfine components in the \amm\, (1,1) line. We note, however, that Figure \ref{fig:hcn_velo} shows intriguing variations in \vlsr\,  between \hcn\, and \amm\, in some regions. One such region is the ridge immediately north of the central cluster, which is aligned north-south. Figure \ref{fig:hcn_velo} b) shows that the \hcn\, line centroids are both red- and blue-shifted around the \amm\, values. In this region, the best agreement between the line centroids is found at the greatest \vlsr\, values. Similar shifts, but with smaller magnitudes, are seen toward the region containing \hcn\, peaks 3a, 3b, 4a, and 4b, shown in Figure \ref{fig:hcn_velo}c. 

\begin{figure*}
\includegraphics[width=0.85\textwidth]{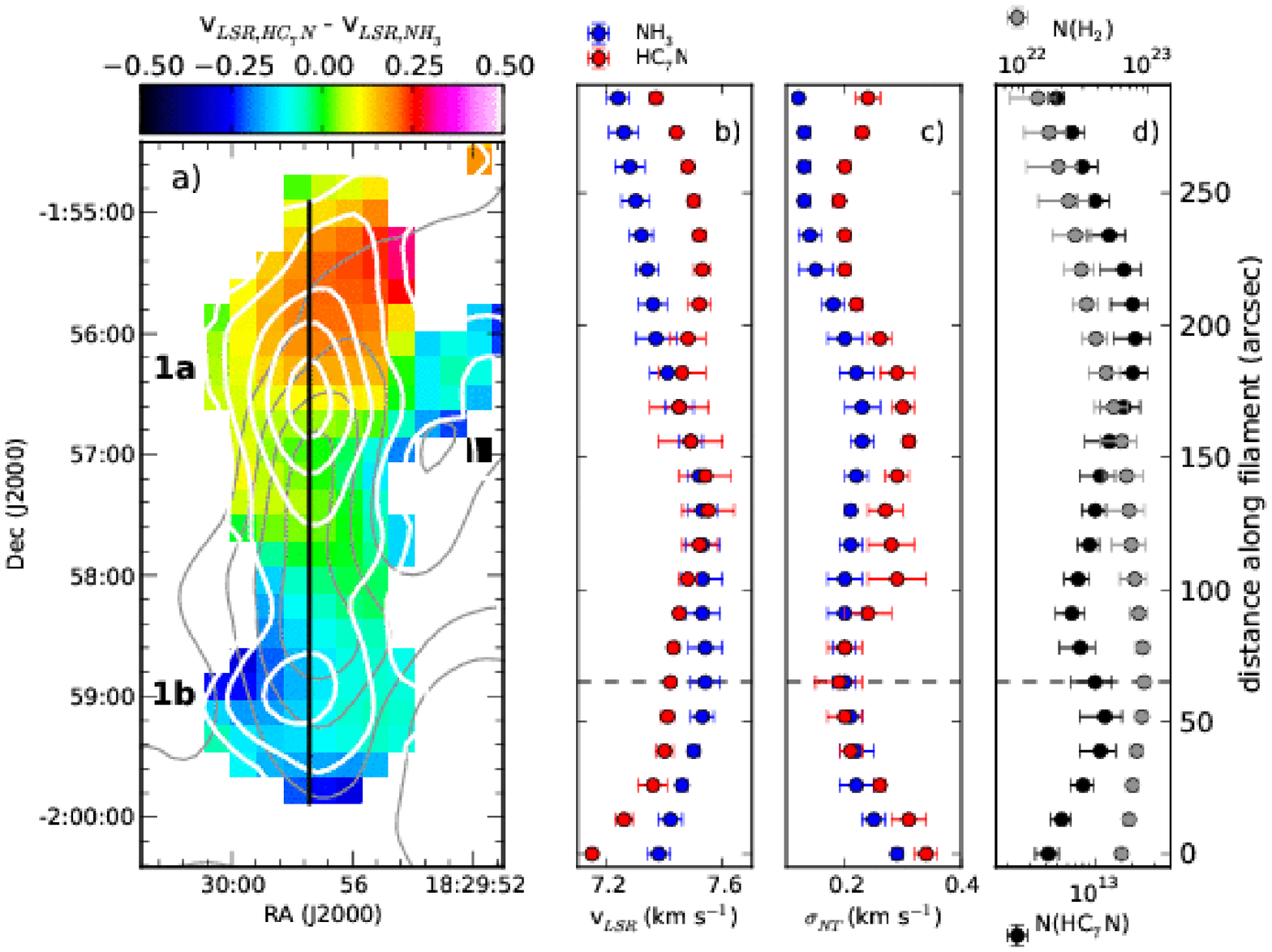}
\caption{a) Map of the difference in \vlsr\,, $v_{LSR, \mbox{\scriptsize{\amm}}}$ - $v_{LSR, \mbox{\scriptsize{\hcn}}}$, between \amm\, and \hcn\, toward the dense filament with \hcn\, peaks labeled 1a and 1b on Figure \ref{fig:hc7n_m0}. White contours show integrated \hcn\, intensity as in Figure \ref{fig:hc7n_m0}. Grey contours show the integrated \amm\, (1,1) intensity. b) \vlsr\, derived from fits of \amm\, (blue symbols) and \hcn\, (red symbols) along the axis shown in black in panel a). Error bars show the $1-\sigma$ spread in values along the perpendicular (R.A.) axis where spectra meet the S/N limit. In this panel and panels c) and d), points are spaced by 13\arcsec. c) The non-thermal line width, $\sigma_{NT}$, for \amm\, (blue symbols) and \hcn\, (red symbols) along the axis shown in black in panel a). The non-thermal line width was calculated by assuming the temperature for both species is the \amm-derived kinetic temperature, $T_K$, at each pixel. Error bars show the $1-\sigma$ spread in values along the perpendicular (R.A.) axis where spectra meet the S/N limit. Toward \hcn\, peak 1b, the non-thermal line widths of both \hcn\, and \amm\, are equal, but \hcn\, traces larger non-thermal motions toward \hcn\, peak 1a. d) \hcn\, (black points) and H$_2$ (grey points) column densities along the axis shown in black in panel a). The x-axes have been scaled to show the same absolute range in column density for each species. Error bars show the $1-\sigma$ spread in values along the perpendicular (R.A.) axis where spectra meet the S/N limit. Dashed grey lines show the location of the $N(\mbox{H}_2)$ peak. \label{fig:ridgeVelo}}
\end{figure*}

In Figure \ref{fig:ridgeVelo}, we show the kinematics traced by \amm\, and \hcn\, toward the \hcn\, 1a and 1b clumps along the axis of the filamentary-like structure traced by the continuum emission in this region. Figure \ref{fig:ridgeVelo}a shows the difference in line centroid between \hcn\, and \amm, $v_{\mbox{\tiny{LSR, \hcn}}} - v_{\mbox{\tiny{LSR, \amm}}}$, toward 1a and 1b. The black vertical line shows the axis along which the difference in line centroid (b), difference in non-thermal line width (c), and column densities of \hcn\, and H$_2$ (d) are plotted, where the plotted points and error bars are the mean and standard deviation of the data in R.A. strips at constant declination. In Figures \ref{fig:ridgeVelo}b, c, and d, we also show the declination of the continuum peak as a dashed grey line. The variation in line centroid between \amm\, and \hcn\, is seen clearly, with \hcn\, tracing gas at higher velocity than \amm\, toward 1a, to \hcn\, tracing gas at lower velocity than \amm\, toward 1b, and a region in the centre (which is not aligned with the continuum peak) where the line centroids agree within the standard deviation. 

Figure \ref{fig:ridgeVelo}c also shows that over much of the north-south ridge, the non-thermal line-widths traced by \hcn\, are significantly greater than those traced by \amm, if we assume equal kinetic temperatures for both species. If we instead determine the temperature required to ensure the non-thermal motions traced by \hcn\, are equal to that of  \amm, the resulting temperature for \hcn\, is $\sim 60$\,K. Given the similarity in critical densities between the \hcn\, and \amm\, lines and the lack of strong, nearby heating sources, this large variation in gas temperature between the species is highly unlikely. In addition, as discussed in \S \ref{sec:chem}, Figure \ref{fig:tkHist} shows that \hcn\, is found primarily toward the colder regions in Serpens South, as traced by \amm, and little \hcn\, emission is seen toward warmer regions. We can thus confidently attribute the difference in non-thermal line width magnitudes to variations in the gas motions each species is tracing. 

\begin{figure*}
\includegraphics[width=0.85\textwidth]{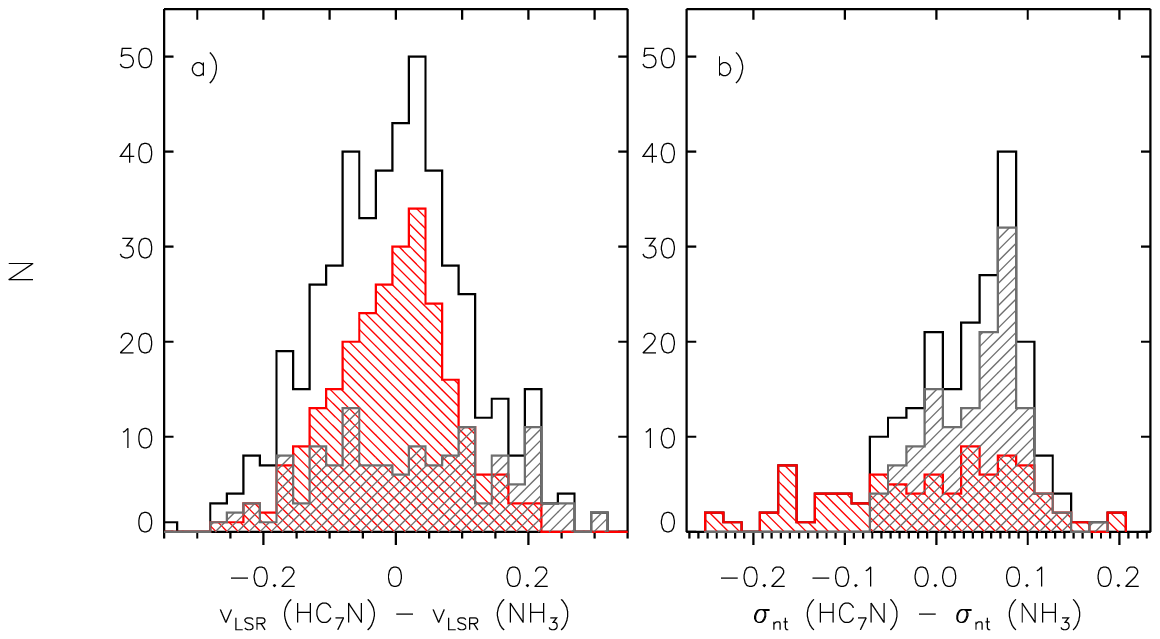}
\caption{Histogram of the difference in a) \vlsr\,, and b) non-thermal line widths, $\sigma_{NT}$, between \hcn\, and \amm\, toward all regions where both \hcn\, and \amm\, are detected above the S/N thresholds (black lines), toward the \hcn\, clumps 1a and 1b only (grey hashed region), and toward all other clumps (red hashed region). Bins are 0.025~\kms\, and 0.02~\kms, respectively. The distributions of the difference in line centroids are consistent with Gaussian distributions for all regions together, as well as when separated into the clumps as described above, but the Gaussian width is substantially larger for clumps 1a and 1b. In contrast, the non-thermal line width difference plot shows a maximum at $\sim 0.07$~\kms\, for all clumps, but the Figure shows this result is dominated by the 1a and 1b clumps, while the other regions show a more smooth distribution between $\sim \pm 0.2$~\kms. Toward \hcn\, clumps 1a and 1b, Figure \ref{fig:ridgeVelo} shows that equal non-thermal line widths between \hcn\, and \amm\, (where a small secondary peak in the distribution can be seen) are localized to the continuum peak just north of the \hcn\, 1b clump. \label{fig:signt} }
\end{figure*}

We show in Figure \ref{fig:signt} histograms of the difference in magnitude between \hcn\, and \amm\, line centroids (a), and non-thermal line widths (b), for all regions where both \amm\, and \hcn\, are detected (black line). For both plots, the grey hatched histogram shows the distributions for the \hcn\, 1a and 1b clumps only, while the red hatched histogram shows the distributions for all other clumps. Note that we are able to derive \vlsr\, toward a greater number of points than $\sigma_{NT}$, which requires an accurate temperature measurement. Over all the \hcn\, clumps, the distribution of the centroid velocity difference between \hcn\, and \amm\, is consistent statistically with being a Gaussian centered at $\Delta \mbox{\vlsr} = 0$~\kms. Fitting the distribution for 1a and 1b separately, we find a small shift in the best-fit Gaussian centre, with $\Delta \mbox{\vlsr} = 0.02$~\kms. There is a larger change, however, in the Gaussian width of the \hcn\, and \amm\, centroid differences between the 1a/1b clumps and all other regions, where we find a Gaussian width of 0.16~\kms\, toward 1a and 1b, but only 0.09~\kms\, toward the rest of the \hcn\, detections. The thermal sound speed, $\sigma_{th} = 0.20$~\kms\, for a gas temperature of 11~K. In general, then, the relative velocities of the two species are therefore subsonic over all the \hcn\, cores, but toward the 1a and 1b cores we see some regions with transsonic variation between the gas velocities traced by \hcn\, and \amm. 

In contrast to the \vlsr\, results, Figure \ref{fig:signt}b shows that the difference in non-thermal line widths between the \hcn\, and \amm\, emission is not Gaussian for the \hcn\, 1a and 1b clumps. In clumps 1a and 1b, the first, smaller peak is centered at $\sim 0$~\kms, which we can see from Figure \ref{fig:ridgeVelo}b is localized near the \hcn\, 1b and continuum peaks. A second, larger peak is seen at $\sigma_{\mbox{\tiny{NT, \hcn}}} - \sigma_{\mbox{\tiny{NT, \amm}}} \sim 0.07$~\kms. For the rest of the clumps, the distribution of the difference in non-thermal line widths is significantly different (red hatched histogram). Here, the distribution of the difference in non-thermal line widths is approximately centered at zero, but with a large standard deviation relative to the results in 1a and 1b. The systematic increase in \hcn\, non-thermal line widths over \amm\, is thus localized to the \hcn\, 1a and 1b region. 

Given the differences seen in line centroids and, in some cases, non-thermal line widths between \hcn\, and \amm, we conclude that the two species are not tracing the same material within Serpens South. In the case of \hcn\, clumps 1a and 1b, the transsonic line widths of the \hcn\, emission suggests it is being emitted in gas that, while still dense, is more strongly influenced by non-thermal motions than traced by \amm. It is possible that \hcn\, is simply tracing more turbulent gas in the outer layers of the filament associated with 1a and 1b, but alternatively, the non-thermal motions in the \hcn\, emission may be dominated by coherent motions, such as infall. If the two species are indeed tracing different material, then abundance variations in the gas are likely present along the line of sight that are not captured by our average abundance calculations above. The mostly likely effect, given the rapid decline in \hcn\, abundance and slow growth of \amm\, abundance at high densities, is that $[\mbox{\amm}]/[\mbox{\hcn}]$ is smallest in the outer, lower density layers around the filaments and cores, but we are unable to probe this further with the data presented here. In addition, the gas temperature where \hcn\, is emitted may be different than that traced by \amm. As discussed previously, however, a small change in $T_{ex}$ will have only a small impact on the \hcn\, column density, as well as on the derived non-thermal line widths. 

\subsubsection{\hcn\, velocity gradients}

\begin{figure*}
\includegraphics[width=0.85\textwidth]{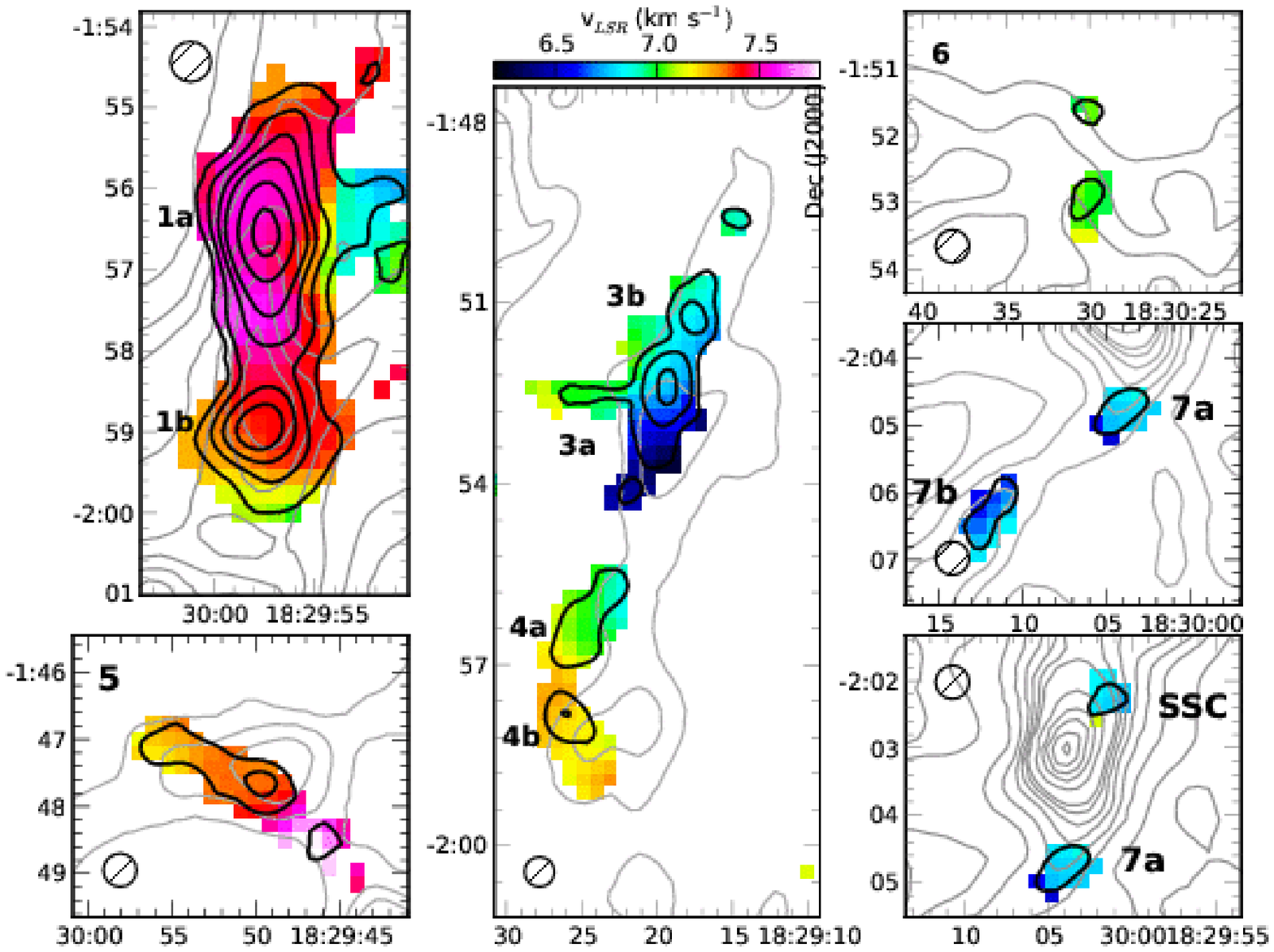}
\caption{\hcn\, line centroid (colour scale) toward seven \hcn\, clumps identified in Figure \ref{fig:hc7n_m0}. We omit \hcn\, clump 2 in the Figure, since the line centroids span an extremely narrow window in velocity ($\sim 0.05$~\kms). Overlaid are 500~\micron\, continuum (grey contours) and  \hcn\, integrated intensity contours (black) as in Figure \ref{fig:hc7n_zoom1}. The 32\arcsec\, FWHM GBT beam at 23~GHz is shown by the hashed circle in each subplot. The Herschel beam at 500~\micron\, is 36\arcsec.   \label{fig:vlsr_zoom}}
\end{figure*}

As Figures \ref{fig:hc7n_zoom1} and \ref{fig:hc7n_zoom2} show, most \hcn\, emission peaks in Serpens South are offset from continuum emission peaks.  \hcn\, clump 2 is an exception, as the \hcn\, emission contours follow closely a small enhancement in the dust continuum emission along a narrow filament extending to the east. Similar patterns of emission from molecular species known to deplete in dense gas, such as CS, have been previously explained as the result of inhomogenous contraction of dense clumps and cores when the emission was correlated with inward motions \citep{tafalla04}. Here, we look at the variation in the \hcn\, line centroids to investigate whether the \hcn\, distribution might arise from a similar process. In Figure \ref{fig:vlsr_zoom}, we show maps of the \hcn\, line centroid toward all the \hcn\, clumps. In regions where the \hcn\, lines could be fit over multiple beams, including \hcn\, clumps 1a, 1b, 3a, 3b, 4a, 4b, and 5, smooth gradients in \vlsr\, are apparent. 

\begin{figure*}
\includegraphics[width=0.85\textwidth]{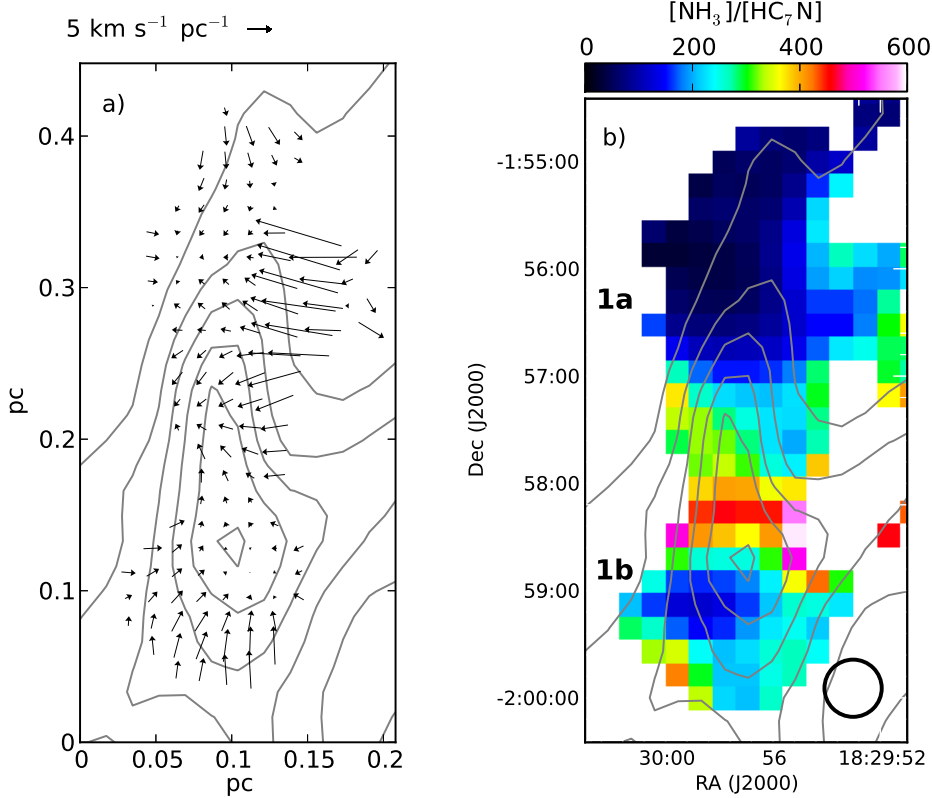}
\caption{Left: Gradient in line-of-sight velocity (\vlsr\,) of \hcn\, emission toward \hcn\, clumps 1a and 1b. In both figures, grey contours show the 1.1~mm continuum emission in intervals of 0.2~Jy~beam$^{-1}$.  Right: Map of the $[\mbox{\amm}]/[\mbox{\hcn}]$ abundance ratio toward the same region as at left. The 32\arcsec\, FWHM beam is shown by the circle in the lower right.\label{fig:hc7nGrad}}
\end{figure*}

We determine the pixel-to-pixel gradient in \vlsr\, using the immediately adjacent pixels in the x- and y-directions (i.e., in R.A. and Decl.), and determine an overall gradient magnitude and direction for each pixel. Toward \hcn\, clumps 1a and 1b, Figure \ref{fig:vlsr_zoom} shows that the \vlsr\, gradients point toward the continuum emission peak along the long axis of the filament. This behavior can be seen more clearly in Figure \ref{fig:hc7nGrad}a, which shows a map of the LOS velocity gradient at each pixel, omitting any pixels immediately adjacent to those masked based on low SNR. The arrows are scaled in length in units of \kms~pc$^{-1}$, given that each pixel is 13\arcsec, or 0.016~pc at 260~pc. Typical uncertainties per pixel are $\sim 0.5 - 0.8$~\kms~pc$^{-1}$ in magnitude, and $\sim  20$\degr\, in angle where gradient magnitudes are $\gtrsim 1$~\kms~pc$^{-1}$ (where gradient magnitudes are smaller, the gradient angle quickly becomes very uncertain). In 1b, the overall gradient is toward the continuum peak from the south- south-east, with a mean magnitude of $2.3 \pm 0.9$~\kms~pc$^{-1}$ and position angle (PA) of $289 \pm 46$\degr\, (east of north). In 1a, there isn't a single overall gradient, but the Figure instead shows coherent gradients from the east, north and west at the northern end of the clump, toward the continuum ridge. Of these, the west-to-east gradient at the filament's north-west edge is the greatest in magnitude, and follows a bend in the larger-scale filamentary structure in Serpens South traced by the continuum emission. Additionally, a gradient across the filament from west to east is clearly seen south of the \hcn\, emission peak. The mean clump gradient is dominated by the west-east motions, with a magnitude of $3.0 \pm 0.9$~\kms~pc$^{-1}$ with PA = $92 \pm 30$\degr. Figure \ref{fig:hc7nGrad}b shows the relative abundance of \amm\, to \hcn\, over the same region, showing that the overall velocity gradients point toward the \hcn\, abundance peaks (or minima in [\amm]/[\hcn], as plotted). The \hcn\, distribution and observed velocity gradients thus suggest that new material is flowing along the larger-scale filament seen in infrared absorption and dust continuum emission, toward the dense region containing \hcn\, clumps 1a and 1b. 

Where they are well-determined (see Figure \ref{fig:vlsr_zoom}), the velocity gradients are remarkably smooth in the other regions, but most do not show similar patterns relative to the dust continuum as seen toward 1a and 1b. Toward 3a, 3b, 4a, and 4b, the \hcn\, emission is largely offset from the continuum and \amm\, emission, and the velocity gradients traced by the \hcn\, line centroid point away from the continuum, along position angles between 40\degr \, to 60\degr\, for 4a, 3a, and 3b, and 110\degr\, for 4b, with mean magnitudes of 1-4~\kms~pc$^{-1}$. A comparison of Figures \ref{fig:hc7n_zoom2} and \ref{fig:vlsr_zoom} shows that the \hcn\, velocity gradients in clumps 3a and 3b are directed toward the peak \amm\, emission, however, which is offset from the continuum peak. The velocity gradient toward \hcn\, clump 5 matches approximately the elongation of the clump with PA $\sim 240$\degr\, and a magnitude of 1.2~\kms~pc$^{-1}$. The PA of this gradient is approximately 30\degr\, offset from the PA of the elongated continuum associated with the \hcn\, clump. Similarly to clumps 1a and 1b, the gradient points toward the \hcn\, abundance peak. 

\subsection{\hcn\, clumps 1a and 1b: Tracing infalling motions}
\label{sec:infall}

\subsubsection{Stability of the filament}

Toward clumps 1a and 1b, the spatial distribution of \hcn, the difference in non-thermal line widths between \hcn\, and \amm, and the velocity gradients seen in the \hcn\, emission may be explained by a pattern of infall of material onto the filament, both along the line of sight and in the plane of the sky. We now examine the stability of the filament containing clumps 1a and 1b to test this possibility. 

We approximate the continuum feature bracketed by \hcn\, clumps 1a and 1b as an isothermal cylinder, and assume the axis of the cylinder to be in the plane of the sky. The continuum emission is elongated, with a long axis length of $\sim 3$\arcmin\, ($0.23$~pc at $d=260$~pc) and an average width of $\sim 1$\arcmin\, ($\sim 0.08$~pc), giving an aspect ratio of $\sim 3$.  This area encloses a region where column densities $N(\mbox{H}_2) \gtrsim 5 \times 10^{22}$~cm$^{-2}$, or half the maximum column density in the filament. For an isothermal cylinder, the radius containing half its mass is similar to the Jeans length at the cylinder central axis \citep{ostriker64}, which ranges from 0.13~pc at a density $n = 10^4$~\cc\, to 0.04~pc at $n = 10^5$~\cc\, at $T = 11$~K, in broad agreement with the width used. The additional structure in the continuum near the filament makes a more rigorous fitting of the feature difficult. We also note that the \hcn\, emission extends $\sim 1.5$\arcmin\, beyond the continuum along the north-south axis, while retaining a similar width, for an aspect ratio closer to 4. 

We estimate the filament mass by summing the H$_2$ column density over a length of $3$\arcmin\, in Dec., to a distance from the continuum emission peak at each decl. of 0.5\arcmin\, along the R.A. axis. We find a total mass $M \sim 31$\,M$_\odot$, and therefore a mass per unit length, $M/L \sim 134$\,M$_\odot$\,pc$^{-1}$. At densities of $n = 10^4$~\cc\, to $n = 10^5$~\cc\, and temperatures $T = 11$~K, the Jeans mass is $\sim 6$~M$_\odot$ to 2~M$_\odot$ \citep{spitzer78}. The calculated mass therefore suggests that the filament should be highly unstable to any density perturbations and consequent fragmentation. 

Assuming no additional support, the maximum $M/L$ for an infinite, isothermal cylinder in equilibrium is $(M/L)_{crit} = 2 k T / (\mu \mbox{\mh} G)$ \citep{ostriker64}. At $T=11$~K, this gives $(M/L)_{crit} = 18$\,M$_\odot$\,pc$^{-1}$, a factor $> 7$ less than the observed $M/L$ ratio. Both \amm\, and \hcn\, line widths exhibit significant contributions from non-thermal motions, with mean velocity dispersions $\sigma \sim 0.2 - 0.25$~\kms. If the non-thermal motions provide support to the filament (i.e. they are not indicative of systematic motions such as infall or outflow), then the adjusted critical line mass is $(M/L)_{crit} = 2 \sigma^2/G \sim 29$~M$_\odot$~pc$^{-1}$, or slightly greater than one fifth of the observed $M/L$ ratio. 

The filament is likely not oriented with the cylinder axis exactly in the plane of the sky. If, instead, the axis is oriented at some angle $\theta$ from the plane of the sky, then the true length $L^\prime = L / \mbox{cos}~\theta$ and the mass per unit length $M/L^\prime = (M / L) \times \mbox{cos}~\theta \sim 134~\mbox{cos}~\theta$~M$_\odot$~pc$^{-1}$. Using the thermal stability criterion, the angle $\theta$ must be greater than $\sim 78$\degr\, for $M/L^\prime \leq (M/L)_{crit}$ to render the filament unstable. This requires the true filament length $L^\prime \sim 1.1$\,pc, which is large relative to the approximate full extent of the Serpens South complex (in projection, $\sim 2.5$~pc measuring along the longest continuous filamentary features). We thus consider this extreme orientation unlikely. In addition, assuming a greater distance of 415~pc to Serpens South (rather than 260~pc), as described in \S \ref{sec:intro}, would increase the total mass in the filament by a factor of 2.5, and the $M/L$ ratio by a factor of 1.6, further bolstering the argument that this feature is unstable. While the unknown orientation of the filament therefore adds uncertainty to the true $(M/L)$ ratio, this analysis shows that the filament is almost certainly unstable to both radial collapse and fragmentation. 

\begin{figure}
\includegraphics[width=0.48\textwidth]{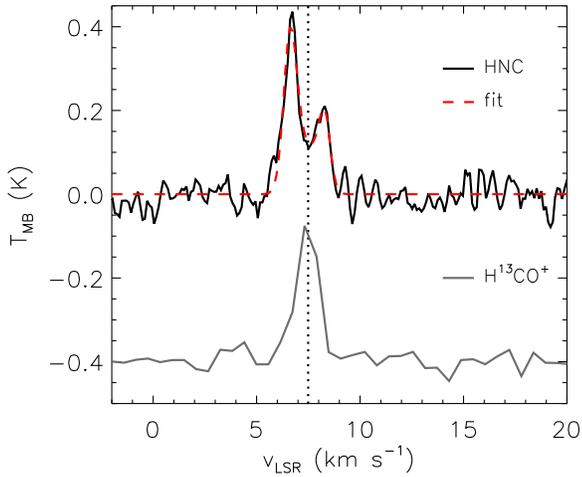}
\caption{Spectrum of HNC $J=1-0$ (black) and H$^{13}$CO$^+$ $J=1-0$ (grey) averaged over the northern filament (H. Kirk, private communication; see \citealt{kirk13}). The H$^{13}$CO$^+$ emission has been offset by $0.4$~K for clarity. The optically thick HNC emission shows a clear blue asymmetry associated with infall in dense regions, with a self-absorption dip that coincides with the peak of the optically thin H$^{13}$CO$^+$ emission and the mean \amm\, (1,1) $v_{\mbox{\tiny LSR}}$ (dotted line). Fitting the HNC data with the `Hill5' infall model \citep[red dashed line;][]{devries05} gives an infall velocity of 0.52~\kms.  \label{fig:hncSpec}}
\end{figure}

\subsubsection{Radial infall}

Given the likely unstable state of the filament, we may expect to see evidence for infall or collapse in the observed gas kinematics. Infall motions in molecular cloud cores are often identified by blue-asymmetric line profiles in optically thick molecular emission lines, where the increasing excitation temperature of the gas toward the core centre results in stronger emission from blue-shifted gas relative to the observer, while self-absorption creates a central dip in the spectral line \citep[e.g.,][]{snell77,zhou92,myers05,devries05}. Since the \hcn\, emission is likely optically thin, we do not see these asymmetric line profiles. Observations of HNC $J=1-0$ emission, however, obtained with the ATNF Mopra telescope (H. Kirk, private communication; see \citealt{kirk13} for details of the observations and data reduction) show this characteristic line profile. We show in Figure \ref{fig:hncSpec} the HNC emission averaged over the northern filament, revealing a blue-peaked line profile with a self-absorption minimum that aligns well with the line centroid for both the optically thin H$^{13}$CO$^+$ $J=1-0$ emission (shown, same dataset) and \amm\, (the mean $v_{\mbox{\tiny{LSR}}}$ over the filament indicated by the dashed line). Using the `Hill5' core infall model described by \citet{devries05}, where the gas excitation temperature linearly increases toward the core centre, and each half of the gas is moving toward the other at a fixed velocity, we find an infall velocity of 0.52~\kms\, (red dashed line in Figure \ref{fig:hncSpec}). This is similar to values found through analysis of self-absorbed spectra toward the filament south of the SSC using the same tracer \citep[$v_{in} = 0.25$~\kms\, and $v_{in} = 0.54$~\kms; ][]{kirk13}. We note, however, that the averaged spectrum has a signal-to-noise ratio of $\sim 15$, whereas the authors recommend a signal-to-noise ratio of at least 30 for high accuracy fits to the infall velocity. Nevertheless, the HNC data show that infall motions are indeed found toward the northern filament, as we expect based on our stability analysis. 

While no asymmetries are seen in the \hcn\, line profiles, infall motions will broaden optically thin emission lines beyond the thermal width \citep{myers05}. Given the strong argument above that \hcn\, must be tracing recently-dense gas, the intriguing consistent increase in the non-thermal line widths of the \hcn\, emission over the \amm\, emission may then be explained by systematic infall of the material traced by \hcn. \citet{arzou13} have shown that supercritical filaments observed with Herschel tend to have larger, supersonic velocity dispersions than subcritical filaments, and suggest that in supercritical filaments, the large velocity dispersions may be driven by gravitational contraction or accretion.  In clumps 1a and 1b, the 0.07~\kms\, difference in non-thermal line width between \amm\, and \hcn\, may then represent a mean infall speed of \hcn\, relative to \amm. This value agrees well with the speeds inferred through observations of infall motions in low mass starless and protostellar cores \citep[e.g.,][]{lee01,williams06}, and with predictions from the \citet{myers13} model of core formation by filament contraction, but is significantly lower than suggested by the HNC modeling. We note that if a portion of the \amm\, non-thermal line width is also due to infall, then our infall velocity estimate is a lower limit for the total infall velocity onto the filament. Additionally, the critical density of the HNC 1-0 line is significantly greater ($\sim 5 \times 10^5$~\cc), and thus is likely tracing infall at higher densities and smaller radii within the filament. More detailed radiative transfer modelling is needed to clarify the relationship between the HNC and \hcn\, line profiles. Here, we are interested in the accretion of material onto the filament, and in the following discussion use the infall speed derived from the \hcn\, line emission, with the caveat that it is likely a lower limit to the true infall speed. 

We can estimate the mass accretion rate onto the filament assuming isotropic radial infall with a velocity $v_{in} = 0.07$~\kms, giving $M_{acc} = v_{in} \rho (2 \pi r L)$. Based on our mass estimate in the filament, $M \sim 31$~M$_\odot$, we find a mean number density $n \sim 5 \times 10^{5}$~\cc. \hcn\, is likely tracing less dense material, however, for two reasons. First, at such high densities \hcn\, is expected to deplete rapidly from the gas, as we discussed above. Second, we showed in \S \ref{sec:chem} that the critical density of the line is substantially less than this value ($n_{cr} \sim 5 \times 10^3$~\cc). The critical density and the mean filament density give lower and upper limits on the density of the gas where \hcn\, is emitted in the region, with the true value likely somewhere in between. Assuming, then, that the density of infalling gas is $\sim 10^4$~\cc, we find an accretion rate of $\sim 5$~M$_\odot$~\mbox{Myr}$^{-1}$ with $L = 0.23$~pc and $r = 0.08$~pc. At $n = 5 \times 10^3$~\cc, we find instead 2.4~M$_\odot$~\mbox{Myr}$^{-1}$, while at $n = 5 \times 10^5$~\cc, the accretion rate is a massive 170~M$_\odot$~\mbox{Myr}$^{-1}$.  

\subsubsection{Infall in the plane of the sky}

Similarly, if we assume the observed \hcn\, velocity gradients also represent gas flow onto the filament, we can estimate the accretion rate in the plane of the sky. This flow could stem from material flowing along the larger filament toward the continuum peak, but could also be the result of the `edge mode' of filamentary collapse that \citet{pon11} showed can occur in filaments of finite length. \citeauthor{pon11} find that the edges of a finite filament can collapse on a shorter timescale than the overall gravitational collapse of an unstable filament, leading to the buildup of dense material at the filament edge. The relative importance of this collapse mode over global gravitational collapse depends on the aspect ratio of the filament. Whether the observed gradients are caused by filamentary flow, or by the gravitational collapse of the filament at its ends, however, in both scenarios the \hcn\, emission reveals newly dense gas accreting onto the filament. The accretion rate along the filament is then $M_{acc,fil} = v_{pos} \rho (\pi r^2)$, where $\rho$ is the infalling gas density, $r$ is the filament radius as above, and $v_{pos} = \nabla v~l$ is the magnitude of the velocity gradient along a length $l$ of the filament in the plane of the sky. 

Here, we look at clumps 1a and 1b separately due to their different velocity centroid patterns. The mean velocity gradient of the 1b clump is 2.3~\kms~pc$^{-1}$ over $\sim 0.1$~pc, north toward the continuum peak. Using the same numbers as above for $n$ and $r$, we find $M_{acc,fil} \sim 3$~M$_\odot~\mbox{Myr}^{-1}$. While clump 1a shows evidence of flow along the filament from the north, the mean velocity gradient is dominated by the west-to-east gradient of 3~\kms~pc$^{-1}$, again over $\sim 0.1$~pc. Because this flow is along the side of the filament rather than along it, we instead estimate the accretion rate as $M_{acc,fil} = v_{pos} \rho (2rL) \sim  6.5$~M$_\odot~\mbox{Myr}^{-1}$, where $L$ is the filament length. Again, if the true filament orientation is at an angle $\theta$ from the vertical rather than in the plane of the sky, then the mass accretion rate will be increased by a factor $(\mbox{cos}~\theta)^{-1}$.  Combining these estimates with the accretion rate derived from our radial infall assumption, we estimate the total accretion rate onto the filament is then $\sim 14$~M$_\odot~\mbox{Myr}^{-1}$, but the uncertainties in this value are relatively large. Given the filament mass, this accretion rate is sufficient to form the filament within $\sim 2$~Myr. This is greater by only a factor of $\sim 2-3$ of the free-fall collapse timescale of a filament assuming a mean density $n = 10^4$~\cc\, and an aspect ratio of 3 \citep{toala12,pon12}, but is significantly longer than the collapse timescale at the density $n = 5 \times 10^5$~\cc\, determined above for the filament. 

\subsubsection{The fate of the filament}

We note that, along the line of sight, the direction of the velocity gradients nearest the Serpens South embedded cluster are away from the cluster. This suggests that the larger filament that the \hcn\, is associated with will not contribute additional material to the ongoing star formation in the SSC. This result is in contrast with similar analysis of the dense filament extending to the south of the SSC, where \citet{kirk13} estimate an accretion rate of $\sim 28$~M$_\odot$~Myr$^{-1}$ onto the central cluster through analysis of the \dia\, velocity gradient along the filament, and a substantial accretion rate lower limit of $\sim 130$~M$_\odot$~Myr$^{-1}$ onto the filament itself through modelling of HNC emission. These accretion rates are similar to the star formation rate in the cluster centre \citep{gutermuth08,maury11}.  

Instead, the already over-dense filament north of the SSC will likely form an additional embedded cluster, with the Jeans calculations above suggesting fragmentation scales of 0.13~pc to 0.04~pc and masses of 6~M$_\odot$ to 2~M$_\odot$. While the continuum emission is smooth, there is some evidence for two separate components within the filament in the \amm\, channel maps. Over several channel maps at lower \vlsr, two peaks in the \amm\, (1,1) emission are seen before the features merge at higher \vlsr, with the brightest emission coincident with the continuum emission peak. The northern \amm\, emission peak overlaps with \hcn\, clump 1a. In contrast, however, both \hcn\, emission peaks persist as separate emission features over all velocities where \hcn\, is detected. The projected distance between the two \amm\, peaks is $\sim 90$\arcsec, or $\sim 0.1$~pc, similar to the Jeans length at $n = 10^4$~\cc. In addition, \citet{bontemps10} identify a Class 0 source coincident with the eastern edge of the continuum feature, suggesting some fragmentation and local collapse may have already taken place. 

\subsection{Other \hcn\, clumps in Serpens South} 
\label{sec:others}

Of the \hcn\, detections in Serpens South, \hcn\, clump 2 is an interesting anomaly. The clump lies within a faint, narrow filament seen in infrared absorption (Figure \ref{fig:hc7n_m0}) and in continuum emission (Figure \ref{fig:hc7n_zoom1}). There is a low-contrast continuum emission enhancement that aligns well with both the \hcn\, emission peak and elongation of the integrated intensity contours, unlike the other \hcn\, clumps. No embedded protostars have been identified. Similarly, faint \amm\, emission is also present, but extends further along the continuum filament in the western direction, and the $[\mbox{\amm}]/[\mbox{\hcn}]$ abundance ratio is the lowest observed in the region, despite the colocation of the \hcn, \amm, and continuum emission (but is similar to those found for 3a, 4a, and 4b). The \hcn\, lines of clump 2 are significantly narrower than those seen toward the other \hcn\, features ($\Delta v = 0.20 \pm 0.04$~\kms, Table \ref{tab:clumpValues}; also see Figure \ref{fig:spec}). Observed offsets between the \hcn\, and \amm\, line centroids are small (Figure \ref{fig:hcn_velo}), and the two species show equal, subsonic non-thermal line widths ($\sim 0.08$~\kms; see Table \ref{tab:clumpValues}). The mass derived from the 1.1~mm continuum emission is $\sim 1.6$~M$_\odot$ within the $\sim 50$\arcsec\, $\times 25$\arcsec\, ($\sim 0.06$~pc $\times 0.03$~pc) \hcn\, contour. These characteristics suggest that the \hcn\, emission is revealing a very young starless core that is forming quiescently within a filament. The model predictions shown in Figure \ref{fig:r12model} suggest that the clump `age', defined here as the time since the gas density reached $n \sim 10^4$~\cc, is approximately a few $\times 10^4$~yr to at most $10^5$~yr. 

The detections of \hcn\, toward other clumps in Serpens South suggests the \hcn\, emission is highlighting regions where newly-dense gas is accreting onto clumps and filaments. With the exceptions of clumps 1a, 1b, 7a, and 7b, the \hcn\, emission is generally associated with relatively low H$_2$ column densities (see Figure \ref{fig:hvsn_column}) and is offset from continuum features (see Figures \ref{fig:hc7n_zoom1} and \ref{fig:hc7n_zoom2}). Apart from clump 2, the \hcn\, line widths are large, with substantial non-thermal components given the cold temperatures of the Serpens South molecular gas. Interestingly, \hcn\, clumps 7a and 7b are associated with the dense southern filament, where asymmetric molecular line profiles show the filament is undergoing substantial ongoing radial infall \citep{kirk13}. Given these results and our interpretation of \hcn\, emission as tracing newly-dense gas, our detections of \hcn\, emission toward the southern filament are expected. We find, however, that the relative abundances of \amm\, to \hcn\, in 7a and 7b are typically high, with a smaller spread in values compared with clumps 1a and 1b, and clump 7b shows the greatest $[\mbox{\amm}]/[\mbox{\hcn}]$ values in the observed region. Given its likely ongoing infall, this suggests that most of the gas accreting onto the southern filament is already at higher densities than the the gas accreting onto the northern filament. 

\section{Conclusions}
\label{sec:summary}

We have detected abundant \hcn\, $J = 21-20$ emission toward multiple locations in the Serpens South cluster-forming region using the K-Band Focal Plane Array at the Robert C. Byrd Green Bank Telescope. \hcn\, is seen primarily toward cold filamentary structures within the region that have yet to form stars, largely avoiding the dense gas associated with small protostellar groups and embedded protostars within the main central cluster of Serpens South.

We calculate the critical density of the \hcn\, $J=21-20$ line, and find that it is similar to \amm. Furthermore, the excitation curve (excitation temperature, $T_{ex}$, vs. density, $n$) of the two species agree well with each other, such that both \hcn\, and \amm\, should be excited in the same regions {\it if} they are both present in significant abundances. We detect \amm\, emission over most of the area mapped in Serpens South. The greatest \amm\, column densities are found along the infrared-dark filaments, but substantial low column density \amm\, is also present between the filaments. Given our high sensitivity observations, it is clear that \hcn\, can only be present in very low abundances in the gas phase toward most of the regions where we find \amm. This abundance difference is likely driven by the different chemical timescales for the formation and destruction of \hcn\, and \amm, as discussed in \S \ref{sec:chem}. 

We show that \hcn\, is not found toward a smooth distribution of H$_2$ column density, in contrast to \amm. Instead, the \hcn\, detections are relatively overdense at both low and high column density (although the distribution is not bimodal). We suggest that this distribution shows there may be two separate regimes in Serpens South in which \hcn\, remains abundant. First, at low $N(\mbox{H}_2)$, where we expect lower gas densities, \hcn\, is present in both chemically and dynamically young regions, such as \hcn\, clump 2, and possibly clumps 3a, 3b, 4a, and 4b. In these regions, we also find very low $[\mbox{\amm}]/[\mbox{\hcn}]$ abundance ratios, suggesting \amm\, has not had sufficient time to form in large abundance, giving a relatively small chemical `age' based on the model in Figure \ref{fig:r12model}. Second, at intermediate to high $N(\mbox{H}_2)$, where gas densities are higher (and the timescale for \hcn\, to deplete from the gas phase is shorter), we argue that \hcn\, is abundant in regions that are actively accreting less-dense material from their surroundings, thereby resetting the timescale for \hcn\, formation and depletion. This interpretation is bolstered by the detailed stability analysis of \hcn\, clumps 1a and 1b in \S \ref{sec:infall}, but can also explain the asymmetric distribution of \hcn\, emission around continuum and \amm\, features in clumps 5, 6, 7a, 7b, and toward the SSC. Toward clumps 1a and 1b, we find consistent variations in the line centroids relative to \amm\, (1,1) emission, as well as systematic increases in the \hcn\, non-thermal line widths, which we argue reveal infall motions onto dense filaments within Serpens South with minimum mass accretion rates of $M \sim 2-5$~M$_\odot$~Myr. Future analysis of the (sub)millimeter continuum and \amm\, emission in Serpens South will probe the stability of the clumps and filaments near the other \hcn\, features. 

We showed that the relative abundances of \amm\, to \hcn\, are similar to previous results where \hcn\, has been detected in nearby star-forming regions, where studies have generally focused on isolated cores in quiescent environments. It is likely, then, that Serpens South is not particularly remarkable in its abundance of \hcn, but instead the serendipitous mapping of \hcn\, simultaneously with \amm\, has allowed us to detect \hcn\, at low abundances in regions where it otherwise may not have been looked for. Our observations, as well as previous mapping observations, suggest that \hcn\, emission peaks are often offset from both \amm\, or continuum emission peaks, although the extended emission overlaps spatially. Searches for the molecule targeted to continuum emission `clumps', without mapping of significant areas, are thus unlikely to have many results. This result extends the known star-forming regions containing significant \hcn\, emission from typically quiescent regions like Taurus and Auriga to more complex, active environments. 

\section*{Acknowledgments}

The authors thank the referee for providing comments that improved the paper. We thank staff at the National Radio Astronomy Observatory and Robert C. Byrd Green Bank Telescope for their help in obtaining and reducing the KFPA data, particularly G. Langston and J. Masters. We also thank F. Heitsch and H. Kirk for helpful discussions, and additionally thank H. Kirk for providing HNC spectra. RF is a Dunlap Fellow at the Dunlap Institute for Astronomy \& Astrophysics, University of Toronto. The Dunlap Institute is funded through an endowment established by the David Dunlap family and the University of Toronto. LM was a summer student at the National Radio Astronomy Observatory, supported by the National Science Foundation Research Experience for Undergraduates program. The National Radio Astronomy Observatory is a facility of the National Science Foundation operated under cooperative agreement by Associated Universities, Inc. RG acknowledges support from NASA ADAP grants NNX11AD14G and NNX13AF08G. This research has made use of data from the Herschel Gould Belt survey (HGBS) project (http://gouldbelt-herschel.cea.fr). The HGBS is a Herschel Key Programme jointly carried out by SPIRE Specialist Astronomy Group 3 (SAG 3), scientists of several institutes in the PACS Consortium (CEA Saclay, INAF-IFSI Rome and INAF-Arcetri, KU Leuven, MPIA Heidelberg), and scientists of the Herschel Science Center (HSC).

\footnotesize{
\bibliographystyle{mn2e}
\bibliography{biblio}
}

\end{document}